\documentclass[iop]{emulateapj}

\usepackage[usenames,dvipsnames]{color}
\usepackage[normalem]{ulem}
\usepackage[colorlinks,urlcolor=blue,citecolor=black,linkcolor=blue]{hyperref}
\usepackage{subfigure}

\newcommand{\e}[1]{\times 10^{#1}}

\newcommand{\msun}{M$_\odot$}

\def\ni {$^{56}$Ni}
\def\nif {$^{58}$Ni}

\begin{document}

\title{Constraints on explosive silicon burning in core-collapse supernovae from measured Ni/Fe ratios}
%\titlerunning{Constraints on explosive silicon burning in core-collapse supernovae}

\author{
A. Jerkstrand\altaffilmark{1},
F.X. Timmes\altaffilmark{2,3},
G. Magkotsios\altaffilmark{3},
S.A. Sim\altaffilmark{1},
C. Fransson\altaffilmark{4},
J. Spyromilio\altaffilmark{5},
B. M\"uller\altaffilmark{6},
A. Heger\altaffilmark{6,7,8},
J. Sollerman\altaffilmark{4},
S.J. Smartt\altaffilmark{1}
}

\altaffiltext{1}{Astrophysics Research Centre, School of Mathematics and Physics, Queen's University Belfast, Belfast BT7 1NN, UK}
\altaffiltext{2}{School of Earth and Space Exploration, Arizona State University, Tempe, AZ}
\altaffiltext{3}{Joint Institute for Nuclear Astrophysics, Notre Dame, IN 46556, USA}
\altaffiltext{4}{The Oskar Klein Centre, Department of Astronomy, Stockholm University, Albanova, 10691 Stockholm, Sweden}
\altaffiltext{5}{ESO, Karl-Schwarzschild-Strasse 2, 85748 Garching, Germany}
\altaffiltext{6}{Monash Centre for Astrophysics, School of Physics and Astronomy, Monash University, VIC 3800, Australia}
\altaffiltext{7}{University of Minnesota, School of Physics and Astronomy, Minneapolis, MN 55455, USA}
\altaffiltext{8}{Shanghai Jiao-Tong University, Department of Physics and Astronomy, Shanghai 200240, P.~R.~China}

\email{a.jerkstrand@qub.ac.uk} 

%\date{}
%\pagerange{\pageref{firstpage}--\pageref{lastpage}} \pubyear{2002}
%\maketitle
%\label{firstpage}

\begin{abstract}

Measurements of explosive nucleosynthesis yields in core-collapse
supernovae provide tests for explosion models. We investigate constraints on explosive conditions derivable from measured
amounts of nickel and iron after radioactive decays using nucleosynthesis networks with parameterized thermodynamic trajectories. The Ni/Fe ratio is for most regimes
dominated by the production ratio of \nif/($^{54}$Fe + \ni), which tends to grow with higher neutron excess and with higher entropy. For SN\,2012ec, a supernova that produced a Ni/Fe ratio of $3.4\pm1.2$ times solar, we find that burning of a fuel with neutron excess $\eta \approx 6\e{-3}$ is required. Unless the progenitor metallicity is over 5 times solar, the only layer in the progenitor with such a neutron
excess is the silicon shell. Supernovae producing large amounts of stable nickel thus suggest that this deep-lying layer can be, at least
partially, ejected in the explosion. We find
that common spherically symmetric models of $M_{\rm ZAMS} \lesssim 13$
\msun~stars exploding with a delay time of less than one second ($M_{\rm cut} < 1.5$ \msun) are able to achieve such silicon-shell ejection. Supernovae that produce solar or sub-solar Ni/Fe ratios, such as SN 1987A, must instead have burnt and ejected only oxygen-shell material, which allows a lower limit to the mass cut to be set. Finally, we find that the extreme Ni/Fe value of 60-75 times solar derived for the Crab cannot be reproduced by any realistic-entropy burning outside the iron core, and neutrino-neutronization obtained in electron-capture models remain the only viable explanation.

% Where in the paper do we actually show that 58Ni/(54Fe + 56Ni) dominate the Ni/Fe ratio? Nowhere, its just stated in words.
% DC 3.4 pm 1.2 in J14
% eta = 6e-3 checked

\end{abstract}


\section{Introduction}

% What are CCSN

Core-collapse supernovae (CCSNe) are the explosive deaths of massive
stars ($M_{\rm ZAMS} \gtrsim 8$ \msun). As the core of the star
collapses to a neutron star or a black hole, a shock wave ejects the mantle and envelope at high
velocities. Nucleosynthesis products from hydrostatic and explosive burning are dispersed into the interstellar medium, providing
a major production channel for the metals in the Universe.
%and CCSNe are
%major contributors to the enrichment of the interstellar medium in the
%early universe.providing

% Current status of our understanding

Comparison of spectral models of standard stellar evolution and explosion
models with observed nebular CCSN spectra show encouraging agreement
\citep[e.g.,][]{Dessart2011, Jerkstrand2012, Jerkstrand2014a}. However, the nature of the explosion
mechanism remains unclear.  One promising scenario is that of delayed
neutrino-driven explosions \citep[see e.g.,][and references therein]{Nordhaus2010, Janka2012, Couch2013, Bruenn2013}. An important test for such models is detailed comparison of explosive nucleosynthesis yields with those inferred from observed nebular-phase spectra. A new method to determine the ratio of the
yields of nickel and iron, after radioactive decays, was presented by 
\citet[][J15 hereafter]{Jerkstrand2015} and applied to several CCSNe.  
Together with literature values, a picture emerged that several
CCSNe show Ni/Fe ratios that are far above the solar ratio. It is of
interest to consider how such  Ni/Fe ratios arise, and whether they
offer constraints on explosion models.

% DC 2015-05 : Nordhaus 2010 

% Siliocon burning

The shock front that travels through the star after the core has
collapsed compresses and heats the overlying layers.  The innermost
layers experience explosive silicon burning which produces iron-group
elements such as \ni, which powers much of the electromagnetic display
of the SN through its decay to $^{56}$Co and $^{56}$Fe. Silicon
burning also produces other isotopes such as $^{57}$Ni, $^{58}$Ni, and $^{44}$Ti,
in amounts that depend on the detailed properties of the progenitor
structure and the thermodynamic conditions. Diagnosis of the yields of
these isotopes can thus provide constraints on the core-collapse
process \citep[e.g.,][]{Hashimoto1989, Thielemann1990, Woosley1991, Thielemann1996, seitenzahl_2014_aa,grefenstette_2014_aa, Perego2015}.

Pioneering calculations of explosive silicon burning were undertaken by
\citet{Truran1966} and \citet{Truran1967}, and the process was
expounded upon by \citet{Woosley1973}. A key property of the burning
in typical core-collapse environments is that it occurs on sufficiently short timescale ($\lesssim 1$ s) that
weak reactions have little time to have a significant influence
\citep{Fowler1964}. The proton and neutron
numbers are therefore preserved, and as most progenitor layers have about equal amounts of protons and neutrons, the ash is $^{56}$Ni rather than $^{56}$Fe. The neutron content of the fuel can be characterized by the neutron excess
$\eta = \left(N_{\rm n} - N_{\rm p}\right)/\left(N_{\rm n} + N_{\rm p}\right)$, where $N_{\rm n}$ and $N_{\rm p}$ are the
number of neutrons and protons, or equivalently by the electron to baryon fraction $Y_e =
\left(1-\eta\right)/2$. For many regimes the burning passes through a
phase of quasi-equilibrium which has weak sensitivity to the initial
composition but a large sensitivity to the value of $\eta$
\citep{Bodansky1968,woosley_1992_aa,hix_1996_aa,hix_1999_aa,meyer_1998_aa,the_1998_aa,Magkotsios2010,Magkotsios2011}. 
There is therefore a direct link between observed iron-group
yields and three fundamental properties of the explosion; the
temperature, the density, and the neutron excess of the fuel. As common 1D progenitor models show the
neutron excess to vary significantly with mass coordinate \citep[e.g.,][WH07 henceforth]{Thielemann1990, Thielemann1996,Woosley2007}, this offers a potential method for constraining which layers are ejected and which are not. Determination of this ``mass cut'' in turn reveals information on the nature of the compact remnant and the explosion mechanism.

In this work we explore what constraints can be derived from measured yields of iron and nickel, as are now
available for several CCSNe (see J15 and references therein). Under most burning conditions, the iron comes predominantly from \ni, and the nickel
comes predominantly from $^{58}$Ni. We focus in particular on SN 2012ec, a Type IIP SN with a progenitor detection \citep{Maund2013}, a well-sampled light curve \citep{Barbarino2014}, and detection of stable nickel lines in the nebular spectrum (J15). %This has motivated our thorough investigation of the burning conditions in a typical core-collapse supernova from a 12-15 $M_\odot$ star.}
In \S \ref{sec:param} we
investigate constraints on the nucleosynthesis obtained from
parameterized thermodynamic trajectories over an extensive peak
temperature - peak density plane. In \S \ref{sec:spherical} we
consider what progenitor layers undergo silicon burning in spherically symmetric
stellar evolution and explosion models.  In \S \ref{sec:discussion} we
discuss the effects of asymmetries and neutrino-processing, as well as implications of the Ni/Fe ratios measured in SN 1987A and the Crab, and in \S
\ref{sec:summary} we summarize our findings.

\section{Parameterized nucleosynthesis models}
\label{sec:param}

We explore models for nickel and iron production using standard, parameterized thermodynamic trajectories. This single-zone approach assumes that a passing shock wave heats material
to a peak temperature $T_{p}$ and compresses it to a
peak density $\rho_{p}$. The material then expands and cools
on a prescribed trajectory until the temperature
and density are reduced to the extent that nuclear reactions
cease (freeze-out). %the assumption of a } evolution . A constant $T^3/\rho$ evolution corresponds to a constant equilibrium radiation entropy. 
Here, we use adiabatic (constant $T^3/\rho$) expansion trajectories 
%\textbf{( equilibrium radiation entropy)
\citep{Arnett1971,Woosley1973}
\begin{equation}
\frac{dT}{dt}    = -\frac{T}{3\tau} \quad
\frac{d\rho}{dt} = - \frac{\rho}{\tau}
\label{eq:ad_ode}
\end{equation}

\begin{equation}
T(t)=T_{p} \exp(-t/3 \tau) \quad
\rho(t) = \rho_{p} \exp(-t/\tau)
\label{eq:Tad_rhoad}
\end{equation}
with a static free-fall timescale for the expanding ejecta \citep{Fowler1964}
\begin{equation}
\tau=(24\pi G\rho_{p})^{-1/2} \approx 446/\rho_{p}^{1/2} \ {\rm s}
\label{eq:AD_timescale}
\end{equation}
% DC the free-fall timescale is derived in Fowler and Hoyle 1964
\citet{Magkotsios2010} demonstrates by comparison with trajectories from several core-collapse simulations that nucleosynthesis yields are generally accurate to within a factor 2 using this treatment. One may also use $\rho(t)$ in the expansion timescale instead of the peak density $\rho_{p}$, see \citet{Magkotsios2010} for examples of such models.
% DC 2015-05 Magkotsios : 44Ti and 56Ni within factor two of actuall CC trajectories

Using the code described in \citet{Magkotsios2010} \citep[see also][]{Magkotsios2011}, we
calculate the mass fractions of nuclear isotopes produced by nuclear burning for different values of $T_p$, $\rho_p$, and initial electron fraction
$Y_e$. We chose peak temperatures and
peak densities spanning the range of $4\times 10^9\leqslant T_{p}
\leqslant 10 \times 10^9$ K, $10^{4} \leqslant \rho_{p} \leqslant
10^{10}$ g cm$^{-3}$. This parameter space covers the conditions
encountered in most CCSN models that produce any significant amounts of Fe or Ni isotopes.
The parameter space is sampled with 121 logaritmically spaced
points, so for any value of $Y_e$ we compute the
final nucleosynthesis at 121$\times$121 points in the ($T_p,\rho_p$) plane. %Using a larger number of sample points does not alter our main results and conclusions.

Our initial composition is the mixture of $^{28}$Si and neutrons that give the
specified $Y_e$; this is achieved by mass fractions $X(^{28}\mbox{Si}) = 1-\eta$ and $X(\mbox{n})=\eta$.
%Our initial composition for any starting ($T_{{\rm p}}$, $\rho_{{\rm p}}$, 
%$Y_e$) triplet is pure $^{28}${Si} for symmetric
%matter ($Y_e =0.5$).  We then add neutrons to make the initial
%composition neutron-rich.  Specifically, we used X($^{28}$Si) = 1 - $\eta$ 
%and X(n) = $\eta$ to set the initial $^{28}$Si and neutron
%mass fractions. 
The choice of initial composition is not important for vast regions of the chosen
thermodynamic parameter space (using e.g. $^{28}$Si and $^{29}$Si to
set $Y_e$ gives the same results) because memory of the initial
composition is quickly erased \citep{Magkotsios2010}.
%DC 2015-01-19, formulas for 28Si and n gives right eta

%The value of the electron fraction $Y_e$ alters the nuclear composition of the final yields, the nucleosynthesis mechanisms for different regions in the peak temperature-density plane, and changes the region's topology. 
For the $Y_e$ parameter, we focus attention on $Y_e =0.490$, 0.495, 0.497,
and 0.499 as these values are representative of different shells in
the pre-supernova progenitor structure (see Sect. \ref{sec:spherical}). Although there are deep-lying layers in
the progenitor with $Y_e$ less than 0.490, this value marks the lower
limit below which \ni~is no longer the major nucleus produced
\citep[e.g.][]{Thielemann1990}, and is therefore of limited interest for this study where we explore SNe with significant $^{56}$Ni production. No
candidate fuel for silicon burning has $Y_e \gtrsim$ 0.499,
and higher values than this are not explored.
% T96 cites also W73 for 0.49 limit

\begin{figure*}[ht]
        \centering
        \begin{subfigure}{
               \includegraphics[width=3.4in]{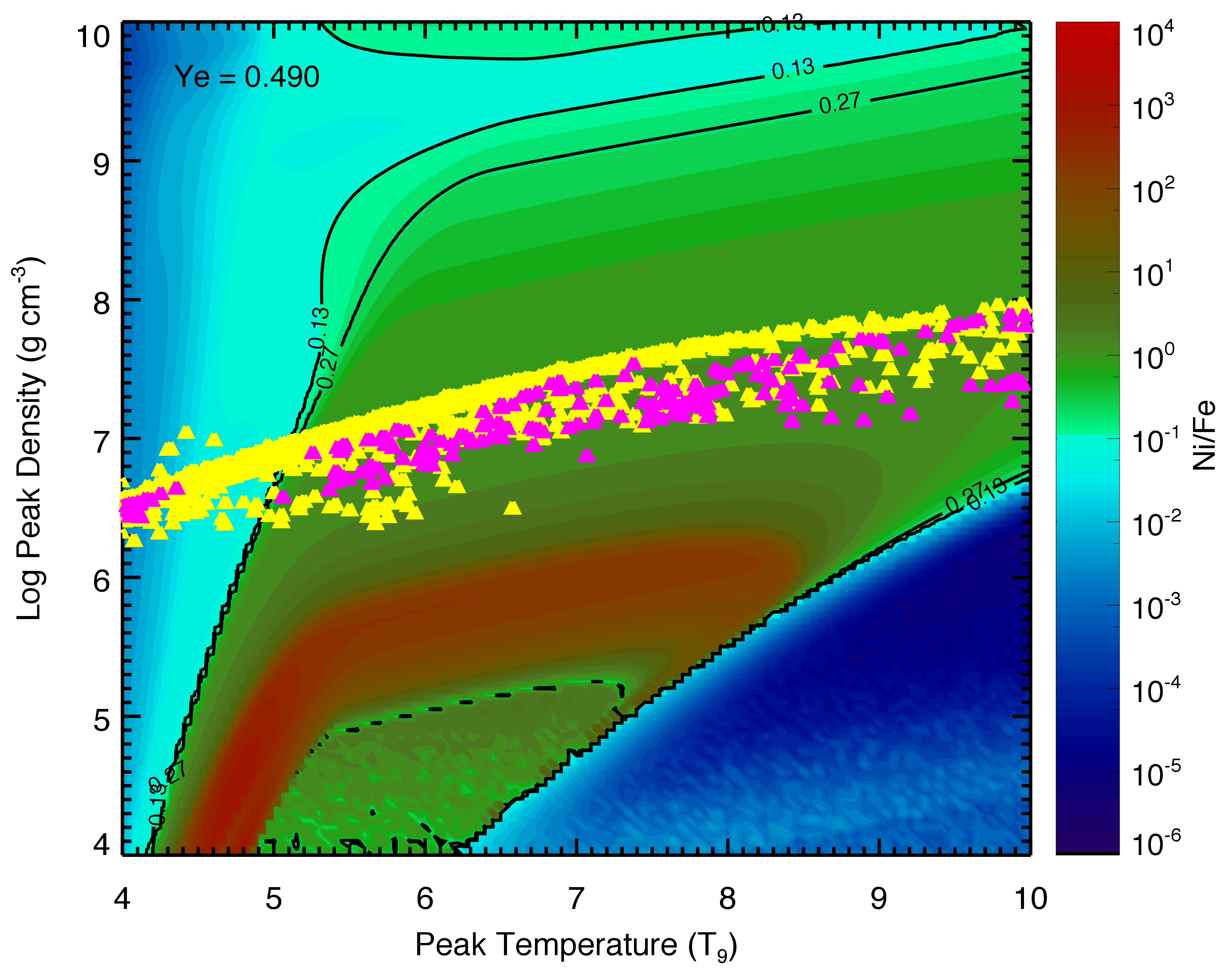}}
                \label{fig:ye0p490}
        \end{subfigure}%
        ~
        \begin{subfigure}{
                \includegraphics[width=3.4in]{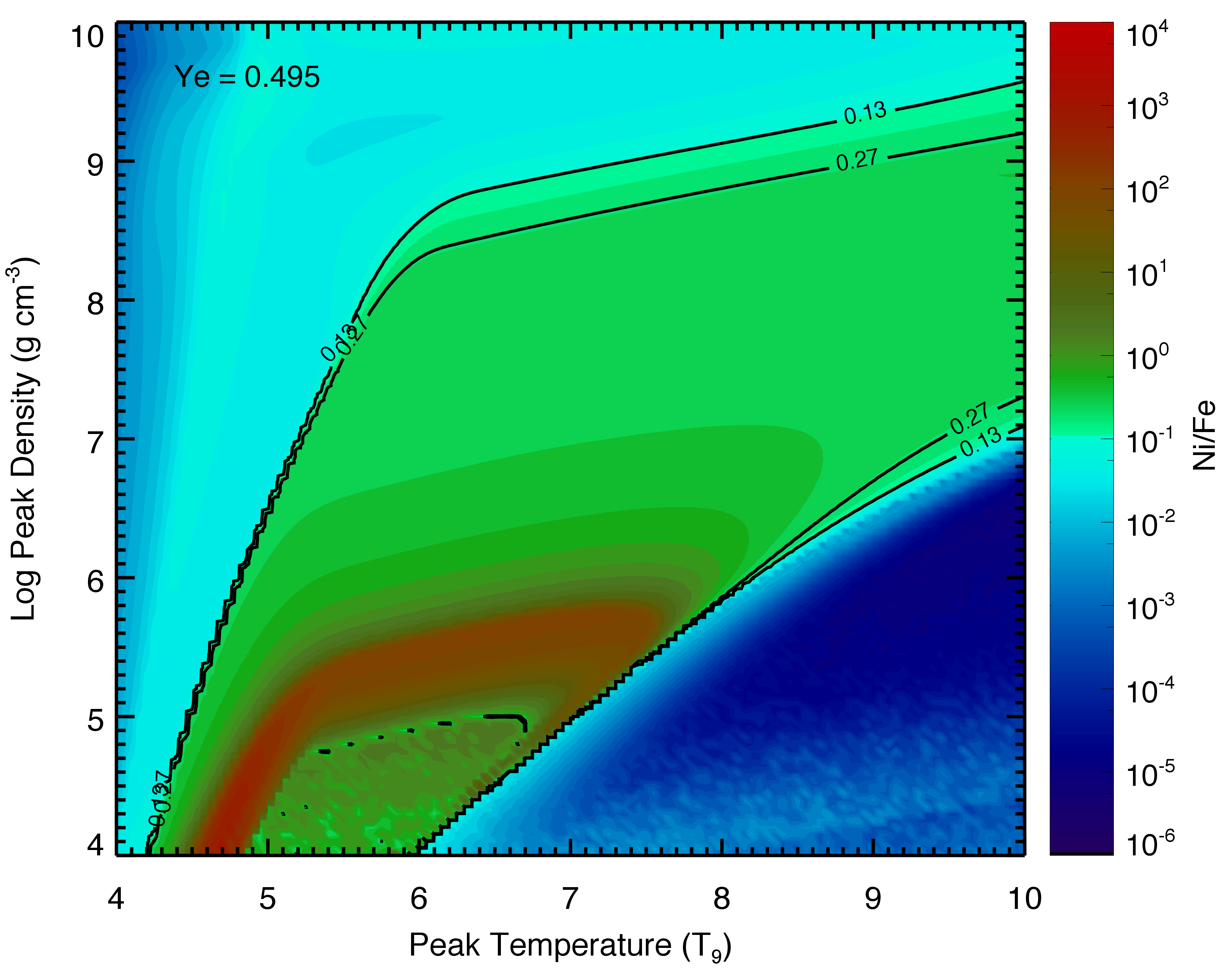}}
                \label{fig:ye0p495}
        \end{subfigure}
        
        \begin{subfigure}{
                \includegraphics[width=3.4in]{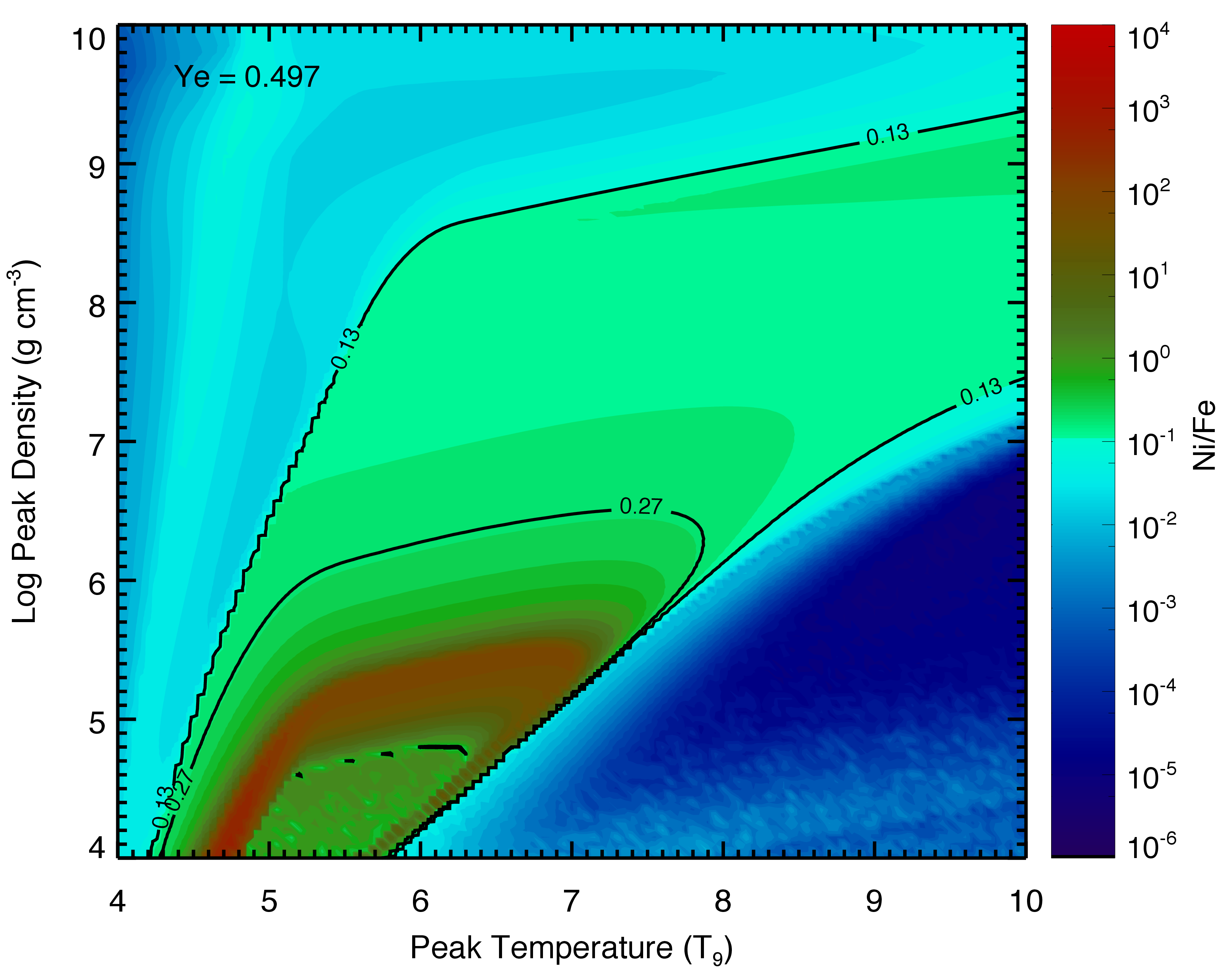}}
                \label{fig:ye0p497}
        \end{subfigure}%
        ~
        \begin{subfigure}{
                \includegraphics[width=3.4in]{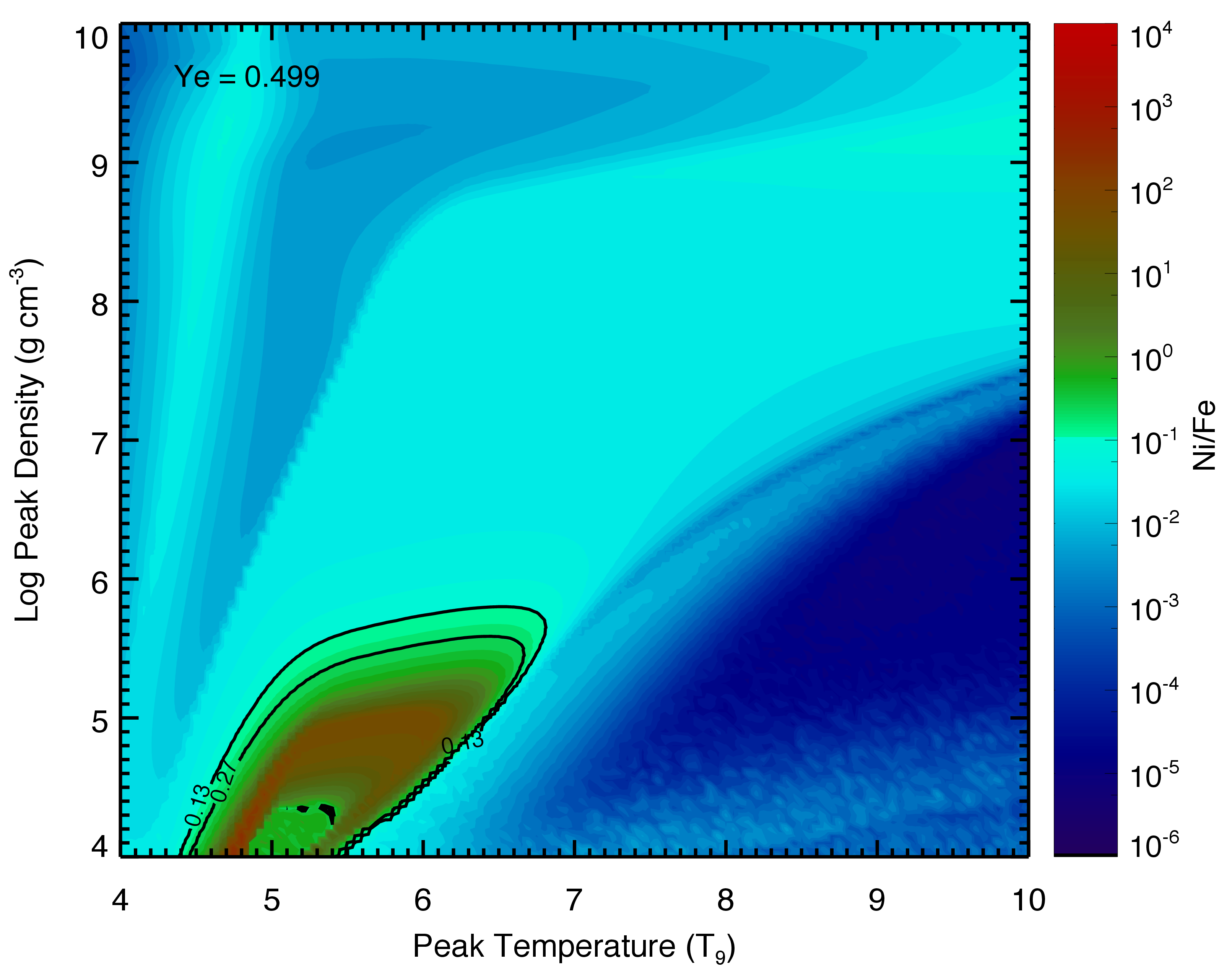}}
                \label{fig:ye0p499}
        \end{subfigure}      
\caption{The Ni/Fe mass ratio (after radioactive decays) in the $\left(T_p,\rho_p\right)$ plane, for an exponential expansion starting with $Y_e$=0.490 (top left), $Y_e$=0.495 (top right), $Y_e$=0.497 (bottom left), and $Y_e$=0.499 (bottom right). On the x-axis $T_9 = T_p/10^9$ K. Black contour lines delineate the value range for SN\,2012ec (0.13 - 0.27). Also plotted in the upper left panel are locations of different mass elements in multidimensional explosion simulations of SN 1987A \citep[yellow,][]{Wongwathanarat2010} and of a 27~\msun~star \citep[pink,][]{Muller2012}.}
\label{fig:yemag}
\end{figure*}

\subsection{Overview Of Results}
\label{sec:overview}

Figure \ref{fig:yemag} shows the resulting Ni/Fe mass ratio (after radioactive decays) in
the ($T_{p}$,$\rho_{p}$) plane for $Y_e$=0.490, 0.495, 0.497,
and 0.499. 
%\sout{The stable nickel abundance sums the contributions from $^{58}$Ni, $^{60}$Ni, $^{61}$Ni, $^{62}$Ni, and $^{64}$Ni, while the stable iron abundance sums the contributions from $^{54}$Fe, $^{55}$Fe, $^{56}$Fe, $^{57}$Fe, $^{58}$Fe, $^{60}$Fe, and $^{56}$Ni.} 
Contour lines are drawn at
Ni/Fe = 0.13 and 0.27, bracketing the range derived for SN\,2012ec in J15 \citep[compare with the solar ratio of 0.056,][]{Lodders2003}.  For Ni/Fe ratios matching SN\,2012ec, the nickel yield is
dominated by \nif, and the iron yield is dominated by \ni~+ $^{54}$Fe,
with \ni~usually the more abundant. We can therefore reach an
understanding of the observed Ni/Fe ratio by looking at the behavior
of these three isotopes.
% DC 2015-05 : At Ye=0.490, 54Fe and 56Ni makes > 90% of iron yield in all regions T9 > 5, except down in the pocket rho < 5.5, T < 7. For nickel, 58Ni dominates same regions (>50%), and all regions T > 6 at 90% level. At Ye = 0.499 the situation is very similar...54Fe and 56Ni make up even more of iron, 578Ni looks similar as at 0.490.

\begin{figure*}[ht]
        \centering
        \begin{subfigure}{
               \includegraphics[width=3.4in]{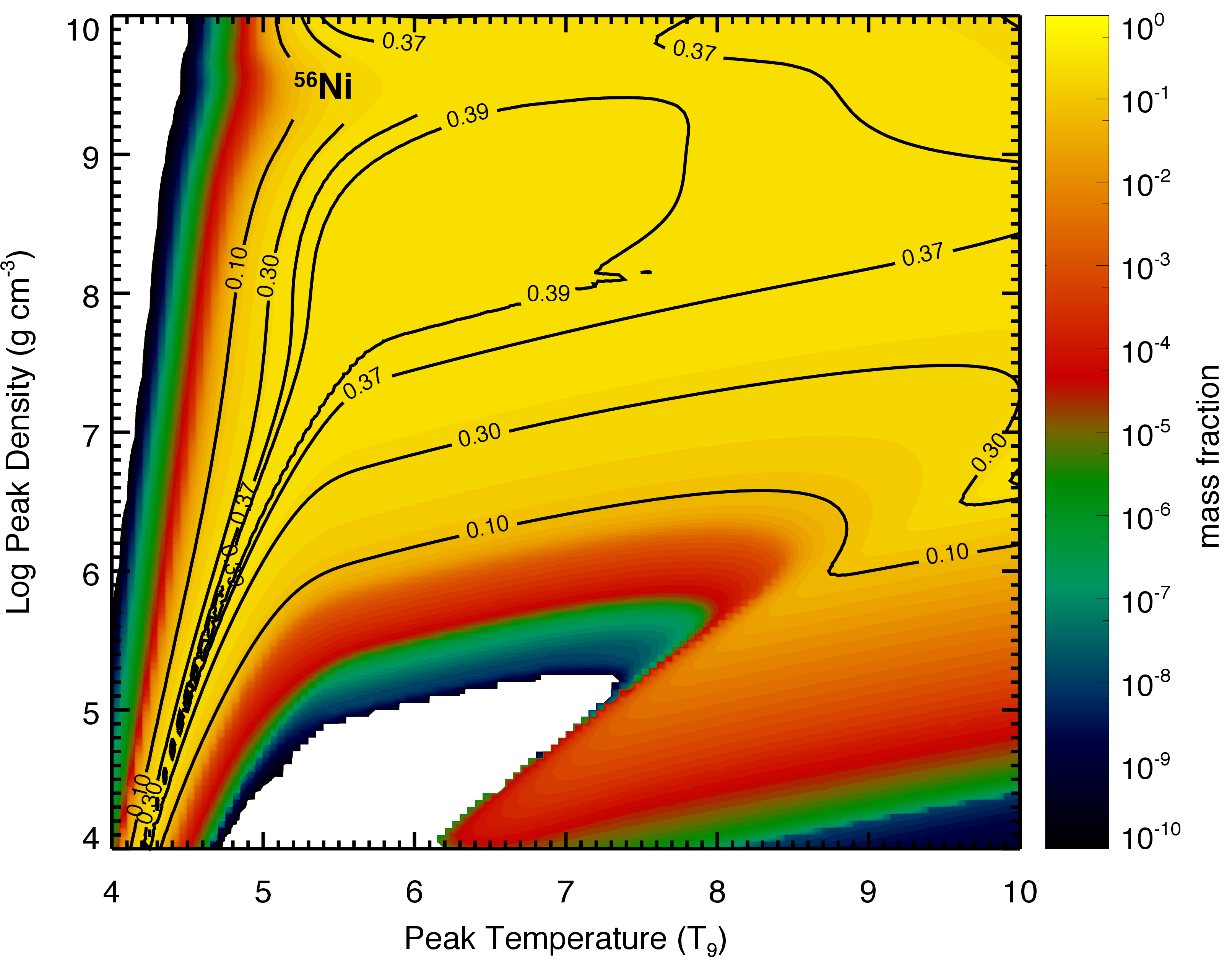}}
                \label{fig:xni56}
        \end{subfigure}
        ~
        %%add desired spacing between images, e. g. ~, \quad, \qquad, \hfill etc.
        %%(or a blank line to force the subfigure onto a new line)
        %\begin{subfigure}{
        %        \includegraphics[width=3.4in]{timmes14_0p490}}
        %        %\includegraphics[width=3.4in]{fig2_ni56}}
        %        \label{fig:yefinal}
        %\end{subfigure}
        \begin{subfigure}{
                 \includegraphics[width=3.4in]{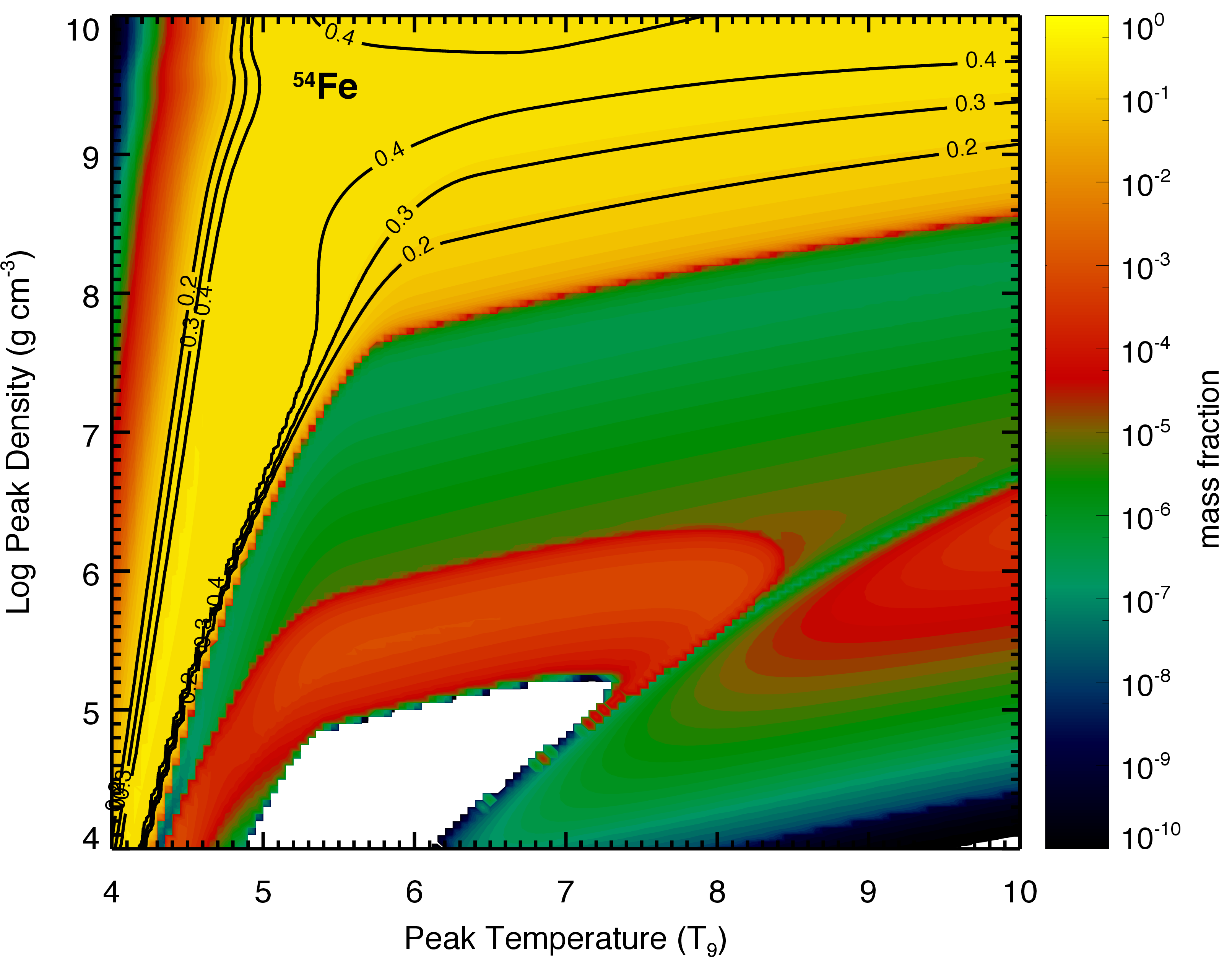}}
                \label{fig:xfe54}
        \end{subfigure}%
        
        \begin{subfigure}{
                \includegraphics[width=3.4in]{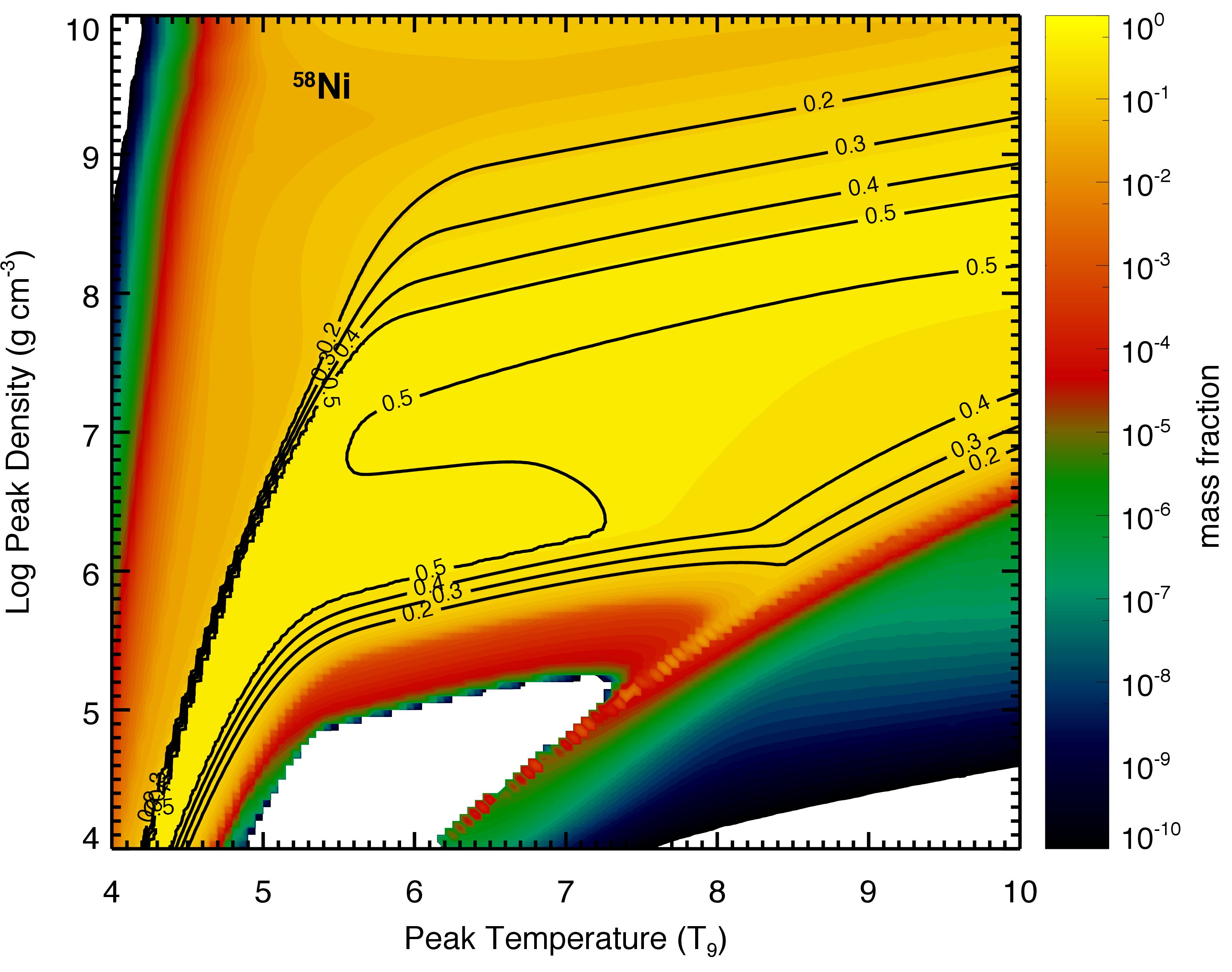}}
                \label{fig:xhe4}
        \end{subfigure}      
        ~
        \begin{subfigure}{
                \includegraphics[width=3.4in]{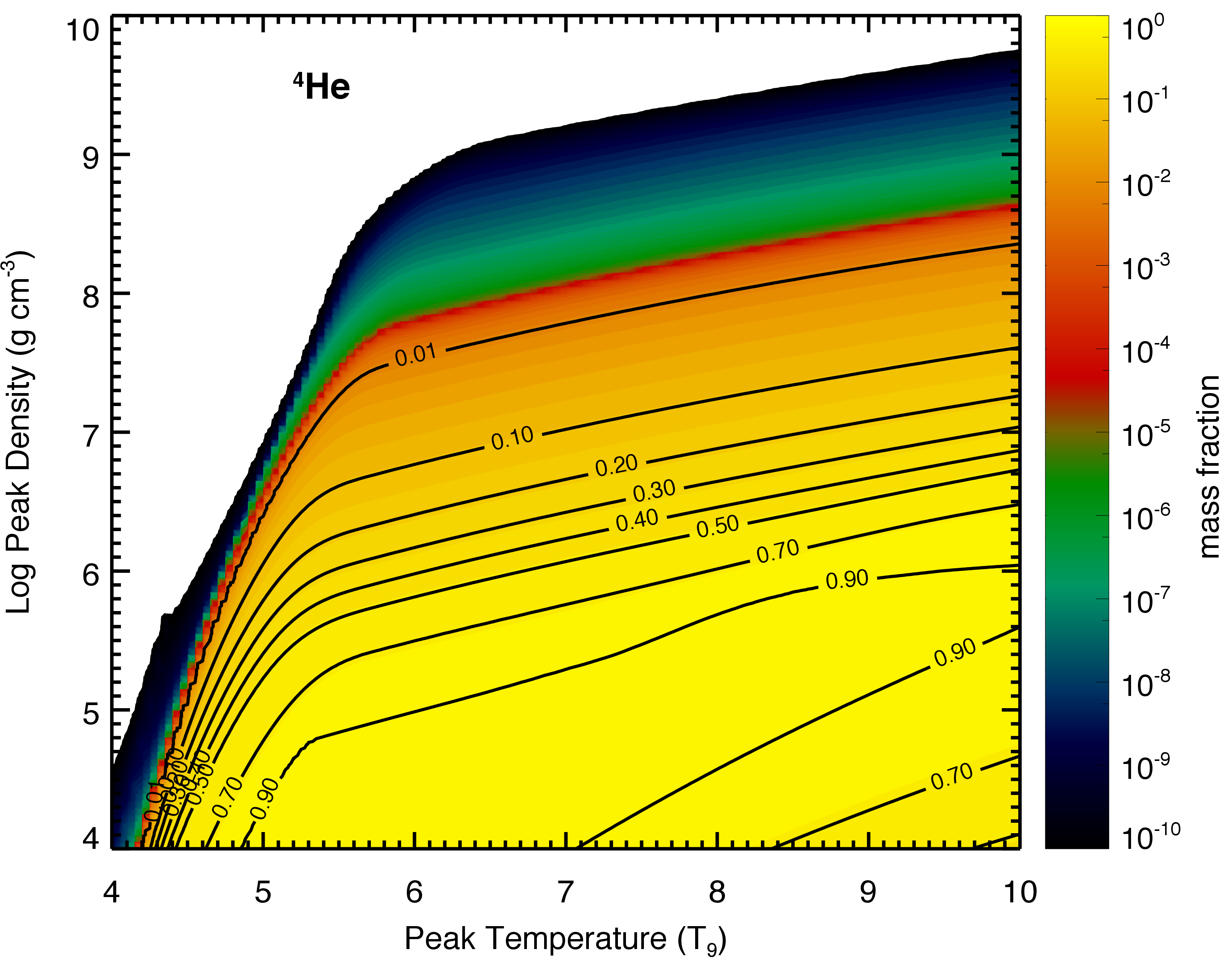}}
                \label{fig:xni58}
        \end{subfigure}%
        %~
        %\begin{subfigure}{
        %        \includegraphics[width=3.4in]{timmes2_0p490}}
        %        \label{fig:entropy}
        %\end{subfigure}      
\caption{The mass fractions of \ni~(top left), $^{54}$Fe (top right), $^{58}$Ni (bottom left), and $^4$He (bottom right), for an initial $Y_e = 0.490$.}
\label{fig:sixpack}
\end{figure*}

% Very rough outline of 56Ni, 54Fe, 58Ni production regimes

Figure \ref{fig:sixpack} shows the mass fractions 
of \ni, $^{54}$Fe, \nif, and $^4$He, for the case $Y_e=0.490$ (other $Y_e$ values
give qualitatively similar trends). 
There are three main production regimes. At $T_9 \lesssim
4.5$ is the incomplete burning regime where little iron-group
production occurs. At $T_9 \gtrsim 4.5, \log \rho_p
\gtrsim 7$ is the complete burning regime with normal
freeze-out\footnote{We adopt here a helium mass fraction $X(^4\mbox{He})\approx0.1$ as the 
dividing line between ``normal'' and ``alpha-rich'' freeze-out regimes.}.
In this regime burning is well described by a single
Quasi-Static Equilibrium (QSE) cluster, in which abundances are largely determined by nuclear Q-values.
At $T_9 \gtrsim 4.5, \log \rho_p \lesssim 7$ is the complete burning
regime with $\alpha$-rich freeze-out, where high production of
$\alpha$-particles depletes iron-group yields. Burning is now
described by two separate QSE clusters, as the triple-alpha
reaction has fallen out of equilibrium. The transition region between the normal
and $\alpha$-rich freeze-out regimes is called the chasm region, here material passes from being dominated by one QSE cluster to two.
At low densities ($\log \rho_p \lesssim 6$) and high temperatures $(T_9 \gtrsim 7)$, $Y_e$ increases during the burning due to weak interactions and mostly
protons and $\alpha$ particles are produced (the so called
$\alpha p$ regime).
For a more refined description of the different burning regimes see \citet{Magkotsios2010}.
% DC 2015-01-16 : Mihalas Mihalas Eq 69.5 gives equilibrium radiation entropy
% S = 4a_RT^3/(3*rho) which is of order 1E8 for typical entropies
% Since R = 8.31E7 in cgs, Sgamma/R is of order unity
% DC 2015-01-19 Thielemann 1990 describes Si burning in three
% fundamental regimes : incomplete, complete with normal freezeout,
% and complete with alpha-rich freezeout. Thw regions are marked in Fig 5; incomplete for T < 5E9 K at low density and 6E9 K at high, alpha-rich starts at rho =1E8 at 5.5 and 5E8 at 10E9 K.

% Rough explanation of why 56Ni, 54Fe, 58Ni change the way they do

\begin{figure}
\includegraphics[width=3.4in]{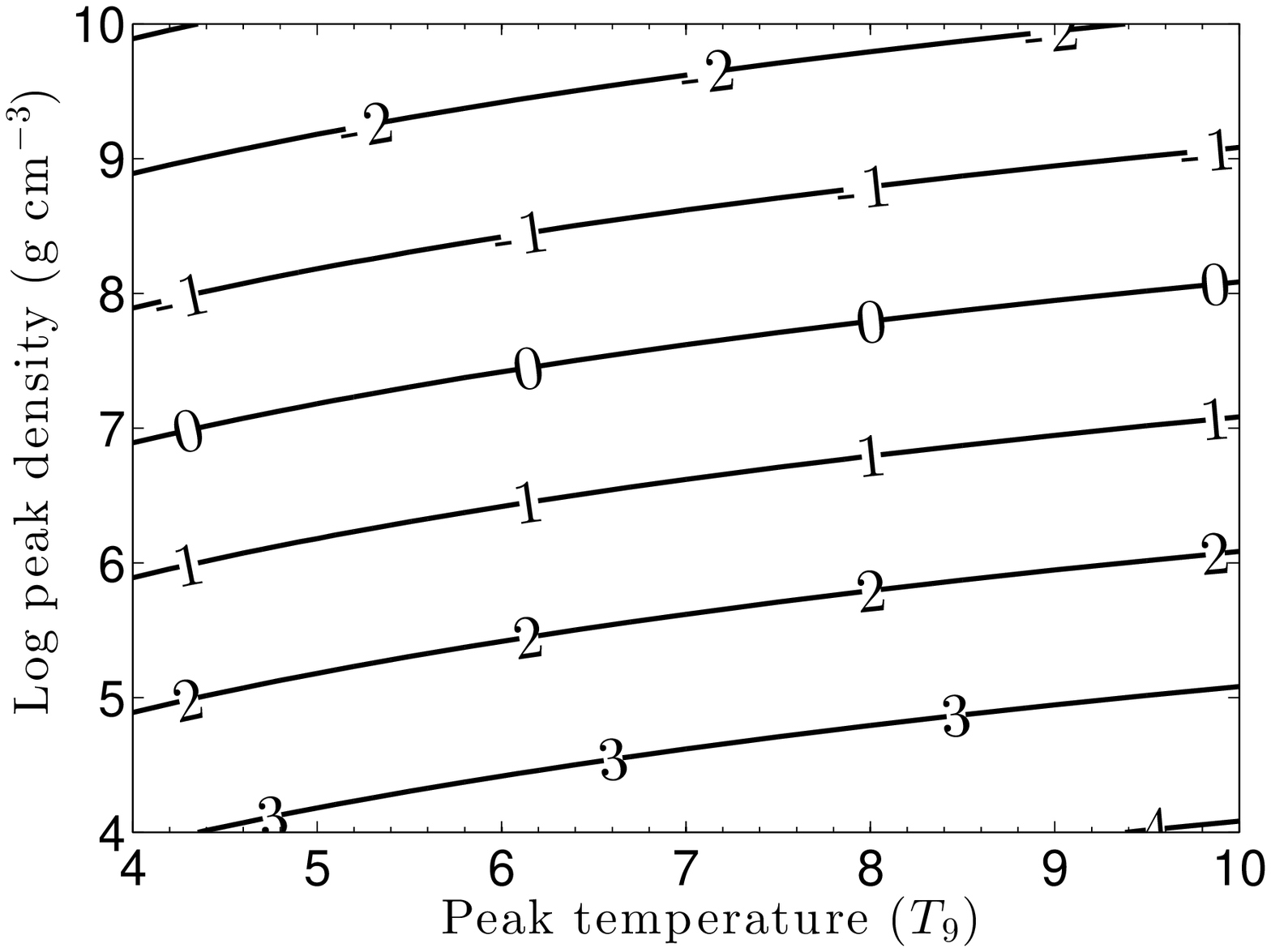}
\caption{Radiation entropy, $\log{S_\gamma/R}$.}
\label{fig:entropy}
\end{figure}

Figure \ref{fig:entropy} shows the radiation entropy $\log\left(S_\gamma/R\right)$, 
where $S_\gamma = 4aT^3/\left(3\rho\right)$ and $R$ is the gas constant. The entropy values will be
of use for later discussion.

\iffalse
The right column of Figure \ref{fig:sixpack} shows the final $Y_e$
(top), the $^4$He mass fraction (middle), and the radiation entropy $\log\left(S_\gamma/R\right)$
(bottom), where $S_\gamma = 4aT^3/\left(3\rho\right)$ and $R$ is the gas constant, for the case $Y_e$=0.490.  Over a large region of the
thermodynamic parameter space the burning occurs under such conditions
that $Y_e$ does not change much from its initial value.  However, at
high densities ($\log \rho_p \gtrsim 9$), $Y_e$ decreases over the
burning time-scale due to electron captures, while at low densities
and high temperatures, $Y_e$ increases due to weak interactions (the so called
$\alpha p$ regime).  In the high-density regime the production of
neutron-rich nuclei will thus be favored as $Y_e$ decreases, whereas
in the low-density regime it will be suppressed as $Y_e$ increases.
\fi

Figure \ref{fig:sixpack} allows a quantitative description of
the \ni, $^{54}$Fe, and \nif~yields, and from them the Ni/Fe $\approx$
\nif/($^{54}$Fe + \ni) ratio. The neutron excess of \ni~is zero,
whereas for $^{54}$Fe and \nif~it is 0.037 and 0.034,
respectively. Thus, as long as these three isotopes dominate the composition, the
neutron excess of the matter must be stored in some combination of
$^{54}$Fe and $^{58}$Ni. Their mass fractions $X$ will then be constrained by
% DC 2015-01-19 : eta(54Fe) = 0.037, eta(58Ni) = 0.034
\begin{equation}
\eta = 0.037 \ X(^{54}{\rm Fe}) + 0.034 \ X(^{58}{\rm Ni})
\enskip .
\label{eq:xx}
% DC 2015-01-19 Formula is exact (somewhat long derivation)
\end{equation}
The maximum mass fractions of these two isotopes are thus 
$X(^{54}{\rm Fe}) < \eta/0.037$ and $X(^{58}{\rm Ni}) < \eta/0.034$.  
The peak plateau values of $^{54}$Fe and $^{58}$Ni production in
Fig.~\ref{fig:sixpack} correspond to the mass fractions that give the
correct neutron excess when that isotope dominates the
composition. %\sout{Note how the $^{54}$Fe abundance increases at very high densities as  $\eta$ then increases during the burning as shown in the final $Y_e$ plot of Fig.~\ref{fig:sixpack}}.

% normal freezeout and alpha-rich freezeout

The neutron excess is predominantly stored in $^{54}$Fe at low entropy and in $^{58}$Ni at high entropy. In the normal freeze-out regime, the Ni/Fe ratio
must therefore increase with increasing entropy as $^{54}$Fe is
replaced by \nif.  When the entropy is large enough, both \ni~and
\nif~are replaced by $\alpha$-particles. For high temperatures, weak
interactions de-neutronize the matter and a decrease in the Ni/Fe
yield follows in the $\alpha$$p$ regime. At lower temperature, where
the original neutron excess is maintained, the Ni/Fe ratio stays high
as \ni~is destroyed more efficiently than \nif~in strong $\alpha$-rich
freeze-out.

\subsection{$Y_e=0.490$}

The Ni/Fe ratio at $Y_e=0.490$ is shown in 
Fig.~\ref{fig:yemag} (top left). In the complete burning regime, there
is a minimum for the ratio at entropy $\log\left(S_\gamma/R\right) \approx  -2$;
at lower entropy the $^{58}$Ni~yield increases due to electron
captures which lower $Y_e$, whereas at higher entropies it increases as
$^{58}$Ni replaces $^{54}$Fe as the main neutron excess storage
nucleus. This growth continues well into the $\alpha$-rich freeze-out
regime ($\log\left(S_\gamma/R\right)\gtrsim +0.5$), but is
quenched in the $\alpha p$-regime. 
The maximum \ni~production is only $\sim$40\% as $^{54}$Fe and $^{58}$Ni are produced in large amounts at these relatively large neutron-rich compositions.% (\sout{the binding energy per nucleon is only 8.388 MeV for \ni}). 

The band bounded by the contour lines for SN\,2012ec lies between the
normal freeze-out regime and the chasm between the normal and
$\alpha$-rich freeze-out regimes \citep[see Fig.~4 of][]{Magkotsios2010}.
%\sout{In this \textbf{band} ($T_9 \gtrsim 5.5$, $\log \rho_{p} > 8$, $\log\left(S_\gamma/R\right) \sim$ -1),} 
Entrance and exit into the
allowed band (along increasing entropy) is driven by switchover between $^{54}$Fe and $^{58}$Ni, while \ni~changes little. Using Eq. \ref{eq:xx} with $\eta=0.02$
gives $X(^{54}{\rm Fe}) < 0.54$ and $X(^{58}{\rm Ni}) < 0.59$. The large plateau
region of maximum \nif~production thus gives a Ni/Fe ratio
$\gtrsim \mathbf{1}$ ($X(^{56}\mbox{Ni}) \lesssim  0.4$), over 18 times the solar value.
% DC 2015-05 : T9 5.5, log rho 8, S/R ~-1
% DC 2015-05 : 0.54 and 0.59 values.

In the incomplete burning regime ($T_9 \lesssim 5$), the
allowed strip between the Ni/Fe = 0.13 and 0.27 contours is very
narrow and it is unlikely that the burning in SN\,2012ec occurred
precisely under these conditions. The same conclusion
can be drawn for the narrow strip delineating the transition into the
$\alpha p$-regime.

\subsection{$Y_e=0.495$}

At $Y_e \geq 0.495$ ($\eta \leq 0.01$), the neutron excess is too small for $^{54}$Fe ($\eta = 0.037$) and $^{58}$Ni ($\eta=0.034$) to dominate the composition, and \ni~($\eta=0$) is the most abundant iron-group nucleus produced.
%However, $^{56}$Ni gradually dominates the final composition (and thus stable iron) for increasing $Y_e$ as the $\eta=0.037$ of $^{54}$Fe is too high for this nucleus to dominate the production at $\eta < 0.02$
% DC 2015-05 56Ni is alwaus more abundance than either 54Fe or 58Ni.
The $Y_e$=0.495 calculation is shown in Fig.~\ref{fig:yemag}, top
right.  The Ni/Fe band allowed by the SN\,2012ec abundance
determinations moves to slightly higher entropies relative to the
$Y_e$=0.490 case.  Note how the Ni/Fe = 0.27 contour is beginning to move
away from the $\alpha$p-regime. The width of
this lower density band is not significant at $Y_e$=0.495, but will
continue to widen and become significant as $Y_e$ increases. The
plateau region of maximum \nif~production has $X(^{58}{\rm Ni})=0.29$, giving about six times solar Ni/Fe.

% DC 2015-05 Eq 4 gives X(58Ni) < 0.29..ratio is about six times solar

\subsection{$Y_e=0.497$}
\label{sec:ye0.497}

The $Y_e$=0.497 calculation is shown in Fig.~\ref{fig:yemag}, bottom left.  This $Y_e$ is interesting because it provides a very large
region in thermodynamic space where the inferred Ni/Fe ratio
of SN\,2012ec is produced.  The allowed Ni/Fe band covers entropies
$-1 < \log\left(S_\gamma/R\right) < +1$, and crosses into the
$\alpha$-rich freeze-out regime.  A transition into the allowed band from
low to high entropy occurs again as the \nif~abundance increases when
the entropy increases past $\log\left(S_\gamma/R\right) \sim -1$
(normal freeze-out). However, the \nif~abundance now levels off to
its maximum value $X(^{58}{\rm Ni})=0.18$ (from Eq. \ref{eq:xx}) before
the upper boundary ($\mbox{Ni/Fe}=0.27$) of the regime is crossed, and gives a Ni/Fe ratio
that lies within the tolerance interval for SN\,2012ec, explaning the
large size of the allowed region at this $Y_e$.

% DC 2015-05  entropy is about -1 to +1.
% DC 2015-05 Eq 4 gives X(58Ni) = 0.18

A qualitative difference to lower $Y_e$ values is that now a part of the
allowed band lies in the $\alpha$-rich freeze-out regime. In
particular, the upper Ni/Fe boundary (0.27) is crossed (with increasing entropy) not as
\nif~increases by too much (as at lower $Y_e$), but by a more rapid
depletion of \ni~compared to \nif~in the
$\alpha$-rich freeze-out regime.

A \nif/\ni~ ratio of 0.2 ratio corresponds to an electron fraction
$Y_e$=0.497 if the abundances of other nuclei are neglegible.  If the freeze-out composition is dominated by these two isotopes, then $Y_e$=0.497 must be the electron fraction of the fuel (assuming $Y_e$ stays constant during the burning). If other isotopes are present at freeze-out, but \ni~still dominates the mass fraction, $Y_e = 0.497$ represents the maximum allowed
electron fraction, as no significant amounts of proton-rich isotopes ($Y_e > 0.5$)
are produced and the other isotopes must contribute zero or positive
neutron excess. However, incomplete burning or a strong $\alpha$-rich
freeze-out may allow also larger $Y_e$ values as \ni~is then not the main nucleus produced.

\subsection{$Y_e=0.499$}

The $Y_e$=0.499 calculation is shown in Fig.~\ref{fig:yemag}, bottom
right. There is no regime at normal freeze-out that produces
enough \nif~to reproduce the SN 2012ec Ni/Fe ratio, as the maximum \nif~fraction $X(^{58}{\rm Ni})=0.06$ (from
Eq. \ref{eq:xx}), and thus Ni/Fe $\lesssim 0.06$ (as long as \ni~dominates).  The only allowed band is restricted to a narrow
region in the $\alpha$-rich freeze-out regime with densities $\log \rho_p <
5.5$.  The allowed band has an entropy of
$\log\left(S_\gamma/R\right) = 1.5 - 1.7$.  At these peak initial
conditions $\alpha$-particles dominate the final composition, and
X($^{56}$Ni) $\lesssim$ 0.3. The allowed band is reached as \ni~is
depleted more strongly than \nif~in the $\alpha$-rich freeze-out.%~during $\alpha$-rich freeze-out. 

% DC 2015-05 entropy 1.5-1.7.
% DC 2015-05 Alpha-particles makes up ~50% of composition in allowed band. Also X(56Ni) <~ 0.3 is correct.

\subsection{Relation to progenitor density and shock velocity}
\label{sec:shockspeeds}

Our $Y_e$ sensitivity study suggests two fundamentally different ways
that a Ni/Fe ratio of $\sim$0.2 can be achieved. The first is a normal
freeze-out burning at low entropies ($-1 \lesssim \log \left(S_\gamma/R\right) \lesssim 0$) of a high neutron excess fuel ($Y_e=0.490-0.497$). The second
is an $\alpha$-rich freeze-out burning at high entropies ($\log
\left(S_\gamma/R\right) \gtrsim 0$) of a lower neutron excess fuel ($Y_e =
0.497 - 0.499$).  At normal freeze-out, the Ni/Fe ratio grows with
entropy because the $^{58}$Ni abundance grows (at the expense of $^{54}$Fe), whereas in an $\alpha$-rich
freeze-out the Ni/Fe ratio grows because \ni~is depleted more
efficiently than \nif~with increasing entropy.% is essentialy constant
%56Ni drops as $\alpha$-aprticles are preferentially produced.

In the non-relativistic, strong shock limit for a radiation-dominated gas
(adiabatic index $\gamma=4/3$), the post-shock conditions are related to pre-shock conditions by \citep[e.g.,][]{Sedov1959, Chevalier1976}
\begin{eqnarray}
\rho_{\rm post} & = & \left( \frac{\gamma+1}{\gamma-1} \right ) \rho_{\rm pre}   = 7 \rho_{\rm pre} \nonumber \\
T_{\rm post} & = & 4300 \ \rho_{\rm pre}^{1/4} \ v_{\rm s}^{1/2}~K,
% DC 2015-01-19 : Density equation in arnett1996 eq 9.9
% DC 2015-01-19 : Chevalier 1976 has (Eq 8) p_2 = 6/7 rho_pre v_sh^2, which with p_2 = 1/3 a T^4 gives T = [18/(7a)]^0.25 rho_pre^1/2 v_sh ^1/2 = 4300 rho_pre^1/4 v_sh^1/2 
% DC 2015-03-06 Janka 2008 has (Eq 1) same Tpost equation constant differ by small factor
\end{eqnarray}
where $v_{\rm s}$ is the shock speed. Thus, a [$T_{\rm post}, \rho_{\rm post}$] pair maps 
onto a [$\rho_{\rm pre}, v_{\rm s}$] pair, as
\begin{eqnarray}
\rho_{\rm pre} & = & \frac{1}{7} \rho_{\rm post} \nonumber \\
v_{\rm s} & = & \left(\frac{T_{\rm post}}{4300 K}\right)^2 \left(\frac{1}{7} \rho_{\rm post}\right)^{-1/2}~\mbox{cm s}^{-1}.
\end{eqnarray}

By equating the post-shock conditions with the peak conditions in the
nucleosynthesis models (i.e., $T_{\rm post}=T_{p}$ and
$\rho_{\rm post} = \rho_{p}$), a determined Ni/Fe ratio constrains the
allowed ranges of $\rho_{\rm pre}$ and $v_{\rm s}$; the values for SN 2012ec are listed
in Table \ref{table:rho1Vs}. More neutron-rich (lower $Y_e$) material requires higher pre-SN densities and slower shock speeds to achieve the same Ni/Fe production.  Note that the $Y_e$=0.499 case requires shock speeds
exceeding $c$/3. Electron capture SNe are capable of achieving shock
speeds around this mark as the shock accelerates down their extremely
steep density gradients \citep{Janka2008}. But SN 2012ec, or any other CCSN producing $M(^{56}\mbox{Ni}) \gg 0.01$ \msun, must arise from a more massive progenitor, and the cores of these are shallower (approximately $\rho \propto r^{-3}$), which allows no significant shock acceleration \citep{Matzner1999}. The initial velocity scale of the shock is limited to $v_{s} \sim \sqrt{2 E/M} \lesssim 0.1c$ from the release of 8.8 MeV per baryon in fusion reactions \citep[see][for a discussion of the energy budget of the explosion]{Scheck2006}. In a self-regulating explosion mechanism, such as the neutrino-driven one, the shock can also never significantly exceed the escape velocity as the energy deposition shuts off as soon as that happens. This gives a constraint $v_{s} \lesssim \sqrt{2GM/R} \lesssim 0.2c$ for $M=1.4$ \msun, and $R=150$ km (typical shock radius prior to explosion). We conclude that the $Y_e = 0.499$ case has no realistic shock scenario associated with it, and can be ruled out as a viable scenario for silicon burning in SN 2012ec.

The $Y_e \leq 0.497$ scenarios all require
physically reasonable progenitor densities and shock speeds. We consider, however, that $Y_e \leq 0.495$ is unlikely because the required entropies are over an order of magnitude lower than encountered in typical core-collapse explosion simulations (a few examples of temperatures and densities obtained in core-collapse simulations are plotted in the upper left panel of Fig.~\ref{fig:yemag}). The much larger regime in thermodynamic space at $Y_e=0.497$ instead overlaps with typical explosion entropies. We therefore conclude that the most natural scenario in which the Ni/Fe ratio of SN\,2012ec is explained is that the SN burnt and ejected a progenitor layer with $Y_e \approx 0.497$.

\begin{table}
\centering
\caption{Constraints on $\rho_{\rm pre}$ and $v_{\rm s}$ in SN 2012ec from application of the strong shock conditions.}
\begin{tabular}{ccc}
\hline
$Y_e$ & $\log \rho_{\rm pre}$ (g cm$^{-3}$)  & $\log v_{\rm s}$ (km s$^{-1}$) \\
\hline
0.490 & 7.2 - 9.2 & 3.2 - 3.6 \\ % 1.4E6 - 2.9E8
0.495 & 7.1 - 8.7 & 3.4 - 3.6 \\% 7.1E5 - 2E8 
0.497 & 5.2 - 8.2 & 3.6 - 4.6 \\
0.499 & 3.2 - 5.0 & 5.0 - 5.5 \\
\hline
\end{tabular}
% Derived from looking at fig 9 in timmes.m
\label{table:rho1Vs}
\end{table}

\section{Stellar evolution and explosion models}
\label{sec:spherical}

So far our analysis has been focused on single-zone models with
parameterized thermodynamic trajectories, allowing us to explore how the Ni/Fe production ratio depends on burning conditions.
In this section, we examine stellar
evolution and explosion models from the literature to explore which progenitors and
explosions can give the required thermodynamic conditions and neutron
excess of the fuel derived in Sect. \ref{sec:param}. This is a challenging step since our understanding of both progenitor evolution and explosion mechanisms is far from complete, but illustrates the role of the Ni/Fe ratio as an important constraint for the testing of both current and future supernova models. We limit
ourselves in this section to spherically symmetric models, providing
some discussion of multidimensional effects in
Sect. \ref{sec:discussion}. We begin by describing the evolution of
the neutron excess during the pre-SN evolution.

\subsection{Neutron excess during pre-SN evolution}
\label{sec:preSNevol}

\begin{figure}
\includegraphics[width=1\linewidth]{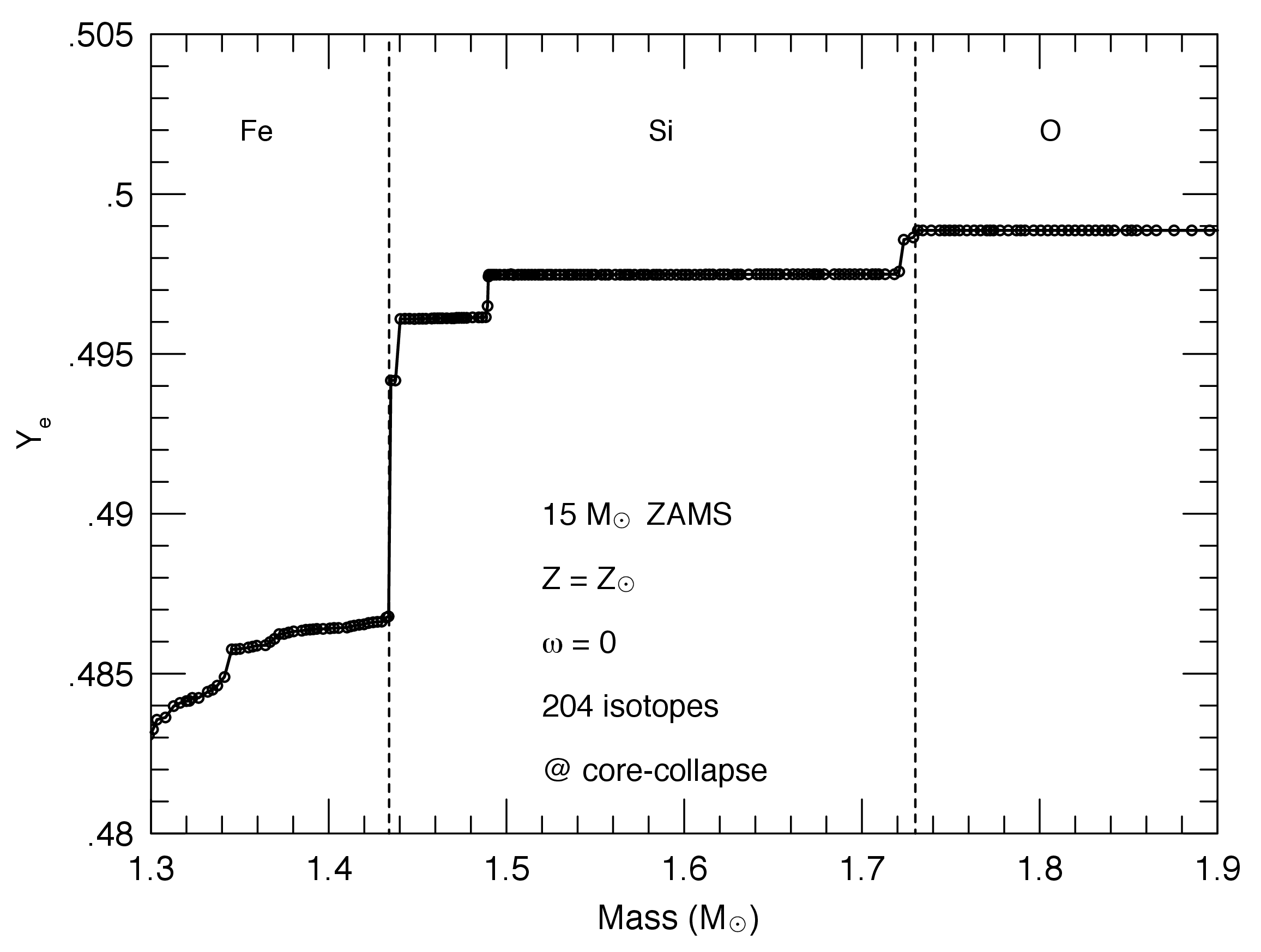}
\caption{$Y_e$ profile at core-collapse of a solar metallicity 15 \msun \ MESA model. The Fe, Si, and O composition layers are labelled,
and open circles mark the locations of grid points in the model.}
\label{fig:yemesa}
\end{figure}

Figure \ref{fig:yemesa} shows the final $Y_e$ profile prior to core-collapse of a solar metallicity, 201 isotope, $M_{\rm ZAMS}$=15 \msun \ model
calculated with MESA, public release version 7315 \citep{paxton_2011_aa,paxton_2013_aa}. 
The boundaries of the He, O, Si, and Fe shells are labeled.  Core
helium burning increases $\eta$ from its initial value close to zero
($Y_e$ = 0.5) through the sequence
$^{14}\mbox{N}(\alpha,\gamma)^{18}\mbox{F}(\beta+)^{18}\mbox{O}(\alpha,\gamma)$\citep[e.g.,][]{couch1972, Arnett1985, Thielemann1985}. The $^{18}$O is burnt further, but in a $\eta$-preserving way. The final neutron excess of the helium burning ashes is therefore the neutron excess of $^{18}$O ($\eta=0.11$) times the mass fraction of material burnt to $^{18}$O (and further), which equals 18/14 times the mass fraction of CNO, as the CNO cycle converts most CNO to $^{14}$N, and the burning of $^{14}$N to $^{18}$O occurs with close to 100\% efficiency in the core. The CNO mass fraction
is about 2/3 of the total metal mass fraction at solar metallicity
\citep{Asplund2009}. Assuming this to hold also for other metallicities, we obtain
\begin{equation}
\eta_{\mbox{He-burn ash}} \approx 10^{-3} \times \left(\frac{Z}{0.014}\right)~.
\label{eq:etaheliumburn}
\end{equation}

Core carbon burning further increases $\eta$ through $^{12}{\rm C}(p,\gamma)^{13}\rm{N}(\beta^+)^{13}{\rm C}(\alpha,n)^{16}{\rm O}$ as well as $^{20}{\rm Ne}(n,\gamma)^{21}{\rm Ne}(p,\gamma)^{22}{\rm Na}(\beta^+)\allowbreak^{22}{\rm Ne}(\alpha,n)\allowbreak^{25}{\rm Mg}(p,\gamma)\allowbreak^{26}{\rm Al}(\beta^+)\allowbreak^{26}{\rm Mg}$ \citep{Thielemann1985}. The increase in $\eta$ is larger for lower-mass cores; about a factor
of two for a solar metallicity 4 \msun~He core but less for more massive ones.  %Consistent with Fig.~\ref{fig:yemesa},
Note that carbon burning will produce a non-zero neutron excess even if $Z=0$. At zero metallicity, \citet{Woosley2002} find $\eta = 1.2\e{-3}$ and $6.8\e{-4}$ for 15\,\msun \ and 25\,\msun~progenitors,
respectively.  As an approximate formula we may take
% DC 2015-05 
\begin{equation}
\eta_{\mbox{C-burn ash}} \approx 10^{-3}\left(1 + \frac{Z}{0.014}\right)~.
\label{eq:etacburn}
\end{equation}

The neutron excess shows no significant changes during core or shell
neon burning. Core oxygen burning increases the neutron excess to
$\eta = 0.01-0.03~(Y_e = 0.485 - 0.495)$ due to electron capture
reactions on $^{33}$S, $^{35}$Cl, $^{31}$P, and $^{32}$S, with higher
values for lower-mass cores \citep{Thielemann1985}.  However, layers
later exposed to explosive silicon burning will, for most progenitors
and mass cuts, have experienced shell oxygen burning rather than core
oxygen burning. The lower density and
higher temperature in shell oxygen burning makes electron captures
less efficient and $\eta$ is smaller \citep{Thielemann1985, Thielemann1990,
  Woosley2002} (see also Fig. \ref{fig:yemesa}).  \citet{Thielemann1990} obtain $\eta \approx$ 0.01 ($Y_e
\approx 0.495$) in this layer, whereas WH07 have $\eta
\approx 4\e{-3}$ ($Y_e \approx 0.498$). In out MESA simultion we obtain $\eta=6\e{-3} (Y_e=0.497)$ in this layer.  During core silicon burning a
large number of electron capture reactions increase $\eta$ to values $\eta \gtrsim 0.06$, ($Y_e \lesssim0.47$, see Fig.~\ref{fig:yemesa}). As for oxygen
burning, shell silicon burning also gives lower neutron excess values,
with \citet{Thielemann1985} obtaining $\eta = 0.04~(Y_e = 0.48)$ in a 8 \msun~He
core, similar to the value obtained in the outer Fe core in our MESA calculation for a 15 \msun~star ($Y_e=0.486$, Fig. \ref{fig:yemesa}).
% DC 2015-05 : Thielemann and Arnett 1985 : No changes in eta during core Ne burning
% DC 2015-05 : Core O burning effect on eta: EC on 33S, 35Cl, 37Ar.
% DC 2015-05 : eta after core O burning: 0.01-0.03 in Thielemann Arnett 1985. Lower mass cores have higher values. Woosley 2002 state eta ~ 0.01 (sec B1) after core O burning.
% DC 2015-05 Thielemann and Arnett 1985 do discuss shell burning, but a bit vague. In their Fig 10 eta after O burn is just 0.005, compared to 0.01-0.03 in Fig 7.
% DC 2015-05 Woosley : T is higher and tho lower in shell burning.
% DC 2015-05 core Si burn gives eta = 0.08 for all MZAMS. Woosley : eta ~ 0.05 at end of core Si burning.
% DC 2015-05 Shell Si burning : Thielemann arnett 1985 have eta ~ 0.04 in their Fig 10.
% Woosley 2002 does not explicitly talk about eta from shell Si burning.

Three distinct zones with very different $\eta$ are thus present in the progenitor; the iron core ($\eta \gtrsim 0.04$), the silicon shell ($\eta \sim 0.01$), and the oxygen shell ($\eta \sim
10^{-3} \times \left(1 + Z/0.014\right)$).  The inferred Ni/Fe ratio in SN
2012ec is produced for $\eta \approx 6\e{-3}$ (Sect. \ref{sec:param}).  There are then two candidate origin sites;
1) The silicon-shell 2) The oxygen-shell in a very high-metallicity progenitor, $Z \gtrsim
0.07$. The second scenario requires five times the solar metallicity, much higher than the measured metallicity of $Z = 0.014-0.025$ (1-1.8 times $Z_\odot$) in the SN 2012ec region \citep{Ramya2007}. We conclude that SN\,2012ec, and likely most other SNe\footnote{Even the highest abundances in the inner few kiloparsecs of spiral galaxies such as M101 are at most twice solar \citep{Bresolin2007} and a value of 5 times solar would be unprecedented.} that produce a
Ni/Fe ratio of several times solar, burnt and ejected part of
the silicon layer.

\subsection{Explosion}

Explosions in spherical symmetry can be modelled
using a thermal bomb \citep[e.g.,][]{Thielemann1996}, a piston-driven event \citep[e.g.,][]{Woosley1995}, or a kinetic bomb \citep[e.g.,][]{chieffi_2013_aa}. \citet{young_2007_aa}
discuss the differences in the yields between thermal bomb and
piston-driven methods. In the specific case of a neutrino-driven mechanism, the explosions can be modelled with self-consistently calculated neutrino luminosities \citep[e.g.,][]{Kitaura2006} or with tuneable ones \citep[e.g.][]{Perego2015}.
% DC 2015-05 T90 uses thermal bomb.
% Woosley (KEPLER) uses piston.
% Chieffi and Limongi uses word "kinetic bomb", they give an initial velocity at the inner boundary.

The explosion mechanism will take some time $t_{\rm delay}$ after initiation of collapse to send the shock wave off. If we specifically define this time as the time at which the shock wave has expanded enough to give a post-shock temperature below the silicon-burning limit, and denote  the radius of the shock at that point as $R_{\rm Si-burn}$, the total mass that will have been burnt is

\begin{equation}
M_{\rm Si-burn} = \int_0^{R_{\rm Si-burn}} \rho(r,t_{\rm delay}) 4 \pi r^2 dr. %= 1.1 M_\odot \left(\frac{\bar{\rho}}{10^7}\right) \left(\frac{E}{10^{51}}\right)
\label{eq:Msiburn}
\end{equation}

A quite accurate estimate of the size $R_{\rm Si-burn}$ of the region that
experiences complete silicon burning can be obtained from the equation \citep{Woosley1988}
\begin{equation}
E = \frac{4\pi}{3} R_{\rm Si-burn}^3 a T_{\rm Si-burn}^4~,
\label{eq:Esiburn}
\end{equation}
Solving for $R_{\rm Si-burn}$ with $T_{\rm Si-burn} = 5\e{9}$ K gives
\begin{equation}
R_{\rm Si-burn} = 3700~\mbox{km} \left(\frac{E}{10^{51}~\mbox{erg}}\right)^{1/3}~.
\label{eq:Rsiburn}
\end{equation}
 For prompt explosions the relevant density profile is that of
the progenitor before any infall, $\rho(r,t_{\rm delay}=0)$. If the explosion is delayed, then
matter has time to accrete and the amount of mass inside $R_{\rm Si-burn}$ will be higher. For example, for $M_{\rm ZAMS} = 13, 15, 20$, and $25$ \msun~progenitors,
\citet[][henceforth T96]{Thielemann1996} find $M_{\rm Si-burn} = 1.42, 1.46, 1.70$, and
1.79 \msun~for $E$=1\,B (1\,B = 1\,Bethe = 10$^{51}$ erg) thermal bomb
explosion and a delay time of zero. For a delay time of one second,
these masses increase to $M_{\rm Si-burn}=1.50, 1.53, 1.77$, and 1.91~\msun.
% DC 2015-01-19 : M_Si-burn in T96 checked table 1 (can also be seen in fig 2)
% DC 2015-01-19 : 1 second delay values checked from Fig 2

The mass cut $M_{\rm cut}$ (the dividing point between matter that
falls onto the compact remnant and matter that is ejected) cannot be
computed ab-initio from thermal bomb or piston-driven models.
Nevertheless, the ejection of even small amounts of iron core material
results in significantly non-solar abundance patterns \citep{Arnett1996}, and one
can therefore argue that most SNe should have their mass cuts above
the iron core (which equals the mass coordinate of the inner edge of
the silicon shell). This mass coordinate is $M_{\rm Fe-core} = 1.18,
1.28, 1.40, 1.61$~\msun~for $M_{\rm ZAMS}=13, 15, 20, 25$~\msun~in the
T96 models.
% DC 2015-05 : M(Fe core) is from table 1 in T96 : 1.18, 1.28, 1.40, 1.61 Msun.
% DC 2015-05 : M(Si

For a given explosion energy $E$ and density profile at the time of shock passage $\rho(r,t_{\rm delay})$ we can thus approximate $M_{\rm Si-burn}$ using Eqs. \ref{eq:Msiburn} and \ref{eq:Rsiburn}. For a given ejected \ni~mass $M_{\rm 56Ni,ejected}$
the mass cut $M_{\rm cut}$ is then
\begin{equation}
M_{\rm cut} = M_{\rm Si-burn} - M_{\rm 56Ni,ejected}~,
\end{equation}
where we have assumed that the mass of the silicon burning ashes is dominated by
\ni. Table \ref{table:masscuts} lists the resulting mass cuts for
different progenitors from the T96 model grid (using $M_{\rm 56Ni, ejected}=0.03$ \msun~as determined for SN 2012ec), and the $Y_e$ value
in the region between $M_{\rm cut}$ and $M_{\rm Si-burn}$ (which
becomes the ejected material that has experienced silicon-burning).

\begin{table}
\centering
\caption{
From the T96 models, the outer mass coordinate for the silicon shell $M_{\rm Si,out}$ (taken as the point where $Y_e$ crosses 0.4985), the inferred mass cuts $M_{\rm cut}$ (assuming zero delay-time, $E$=1\,B, and a
Si-burning ash mass of 0.03 \msun), 
and the $Y_e$ values in the silicon-burning region for those mass cuts.
}
\begin{tabular}{cccc}
\hline
$M_{\rm ZAMS}$  & $M_{\rm Si,out}$ & $M_{\rm cut}$  &  $Y_e$ \\
\hline                       
13        & 1.49            &  1.39            & 0.491 \\
15        & 1.37            &  1.43            & 0.499 \\
20        & 1.63            &  1.67            & 0.499 \\
25        & 1.64            &  1.79            & 0.499 \\
\hline
\end{tabular}
\label{table:masscuts}
% M_Si-out from where Ye makes its first drop in Fig 2 in T97
% DC 2015-01-19 : 1.49, 1.37, 1.63, 1.64 for M_Si,out
% DC 2015-01-19 : M_Siburn from table 1 (1.42, 1.46, 1.70, 1.79)
% DC 2015-01-19 : Mcut = 1.39, 1.43, 1.67, 1.76 from those
% DC 2015-01-19 : Ye = 0.491, 0.499, 0.499, 0.499
\end{table}

\begin{figure}
\includegraphics[width=1\linewidth]{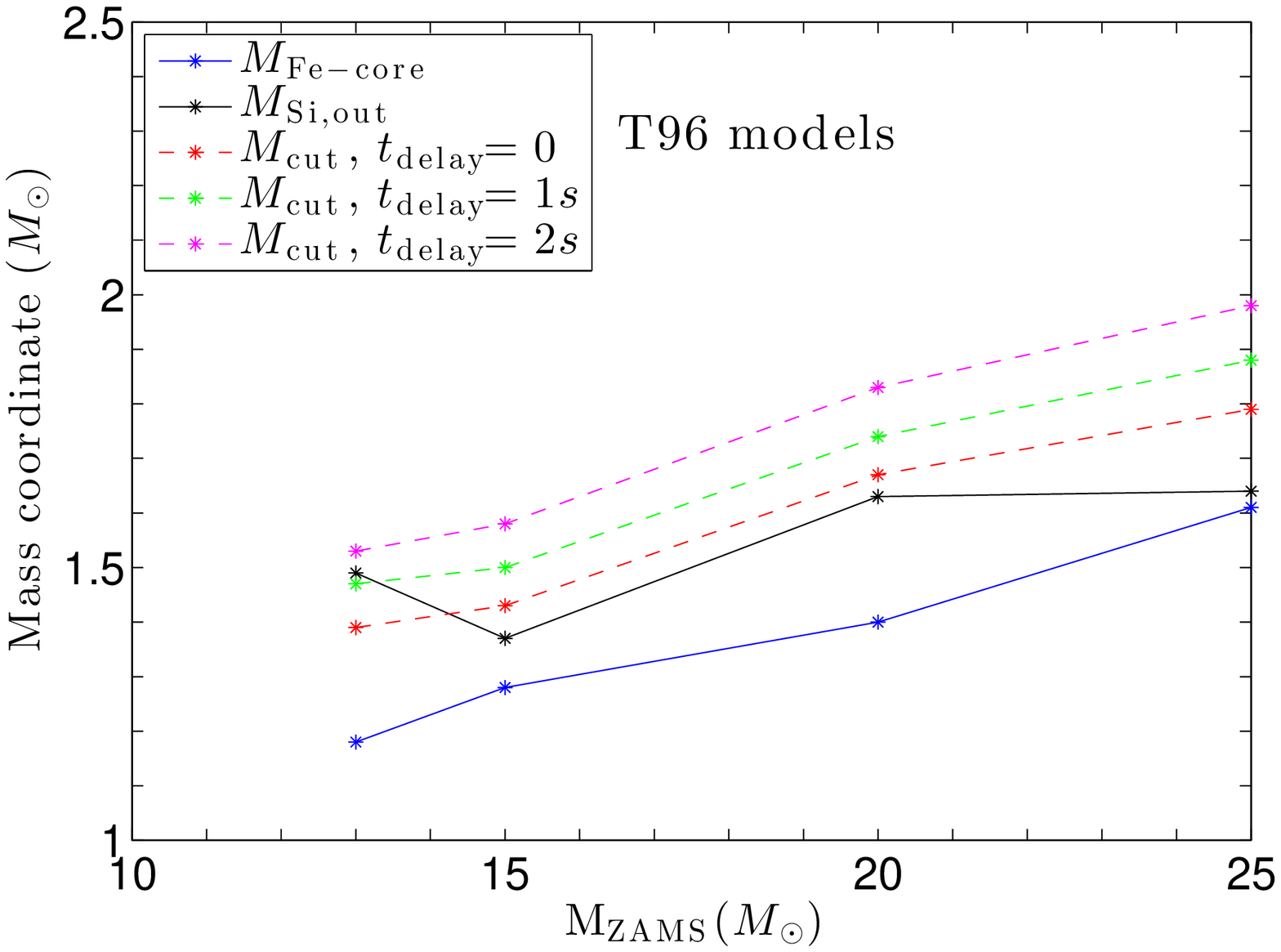} % made by progenitoreta.m (fig 1)
\includegraphics[width=1\linewidth]{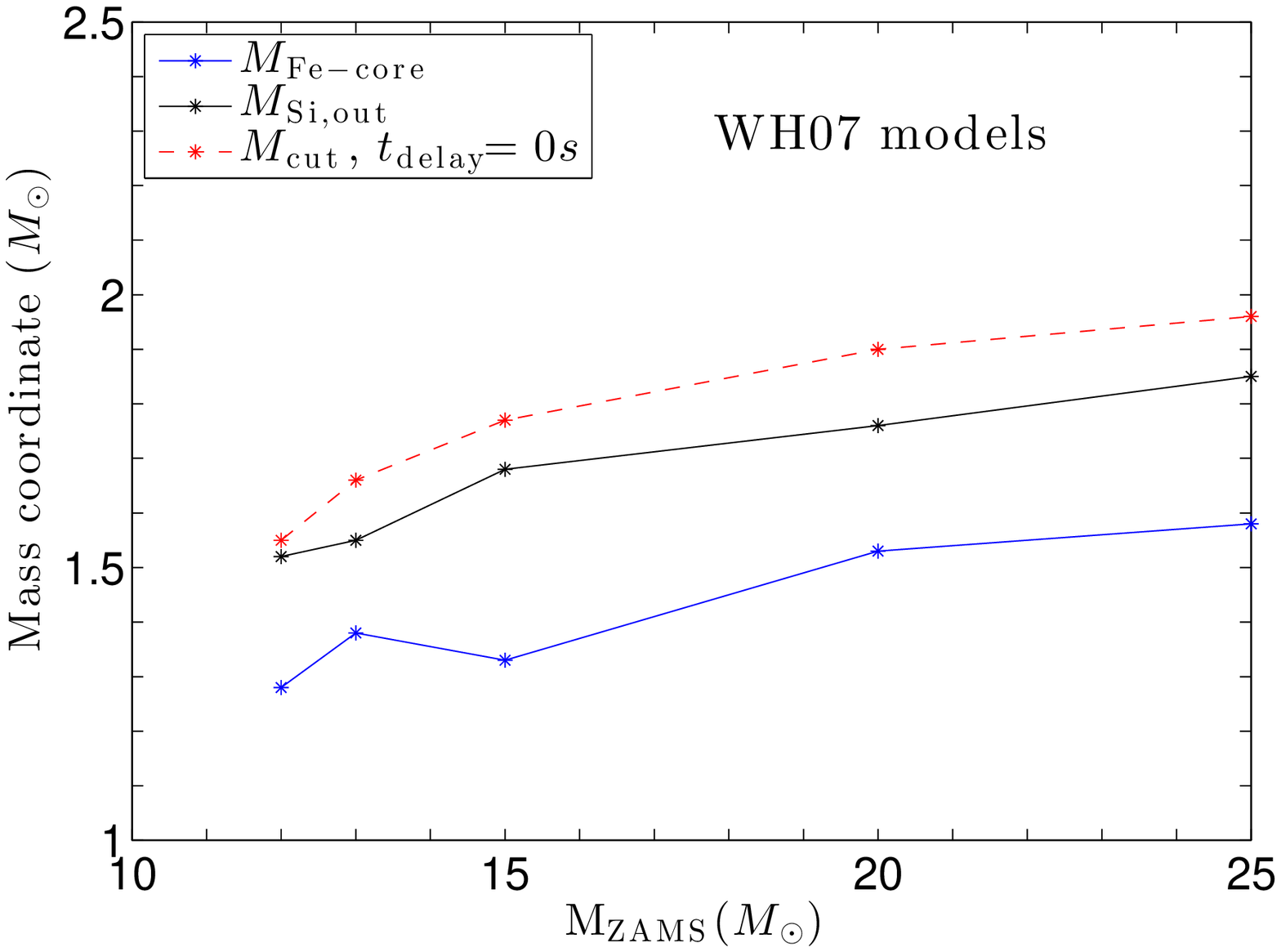} % made by progenitoreta.m (fig 2)
\caption{The mass cuts for $E$=1\,B and $M$($^{56}$Ni)=0.03 \msun, compared to the
outer edge of the silicon shell $M_{\rm Si,out}$ (black, solid), from
the T96 models (top) and the WH07 models (bottom). In the T96 models,
ejection of silicon-shell fuel occurs only for the $M_{\rm ZAMS}=13$
\msun~progenitor. In the WH07 models, silicon-shell material is not
ejected for any progenitor mass. }
\label{fig:MZAMSgrid2}
\end{figure}

Figure \ref{fig:MZAMSgrid2} (top) shows the relevant quantities from
the T96 grid, also plotting $M_{\rm cut}$ for non-zero delay times. For $E \geq 1$ B, only the combination of a $M_{\rm ZAMS}=13$
\msun~progenitor and a delay time of less than one second ejects any
silicon-layer material. The silicon-shell $Y_e$ in this particular model ($Y_e=0.491$) is too low for SN\,2012ec, but the value of $Y_e$ in the silicon shell depends on details in the stellar evolution model. For larger $M_{\rm ZAMS}$ and/or longer delay
times, the mass cut lies above the outer edge of the silicon shell, so
only oxygen-shell material with $Y_e=0.499$ is ejected.

Figure \ref{fig:MZAMSgrid2}
(bottom) shows the same quantities for the WH07 models. These models
%have slightly higher explosion energies of 1.2 B, are piston-driven,
have slightly denser cores for a given $M_{\rm ZAMS}$. The
trends are, however, similar. The WH07 models show the same behavior of the
outer edge of the silicon shell approaching the mass cut for low
$M_{\rm ZAMS}$; in this grid they converge around $M_{\rm ZAMS}=12$\,\msun. The mass cut lies
just outside the silicon shell here rather than just inside as in the T96
models, but small changes in delay time and/or explosion energy will
move the boundary.
% CARE we dont care about actual explosion, only look at rho(r) here

\begin{figure}
\includegraphics[width=1\linewidth]{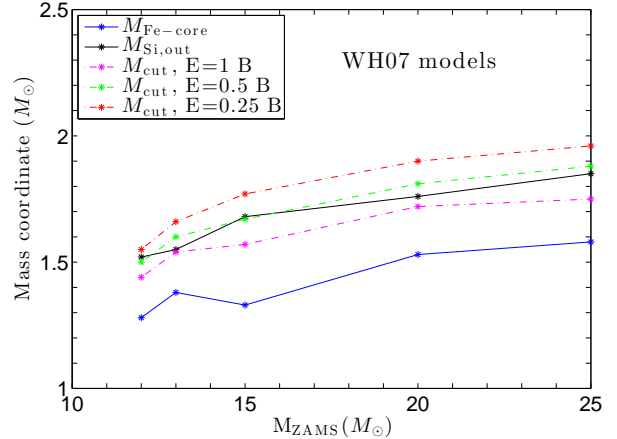} % made with progenitoreta.m
\caption{The position of $M_{\rm cut}$ for different values of $E$, using the WH07 models.}
\label{fig:varyE}
\end{figure}

Using either the T96 or WH07 model grids gives a consistent picture that ejecting the silicon layer at $M_{\rm
  ZAMS}> 13$ \msun~(and synthesizing $M(^{56}\mbox{Ni}) \sim 0.03$ \msun) would require an
explosion energy smaller than 1\,B. Figure \ref{fig:varyE} shows how
$M_{\rm Si-burn}$ varies if $E$ is smaller than 1\,B. At $E$ = 0.5\,B,
ejection of silicon-layers could occur up to $M_{\rm ZAMS} = 15$
\msun, but at larger $M_{\rm ZAMS}$ still only oxygen layers are
ejected. Going to $E$ = 0.25\,B does allow for ejection of
silicon-layers at higher $M_{\rm ZAMS}$. However, the expansion
velocities scale as $V \propto \sqrt{E/M_{\rm ejecta}}$, and for such
a low explosion energy we would expect at least a factor two more
narrow lines than usual. %In addition =, $M_{\rm ejecta}$ will be higher, reducing $V$ further, but this is a second order effect (the ejecta mass goes from 9 to 14 \msun~when $M_{\rm ZAMS}$ goes from 12 to 25 \msun).  
In SN\,2012ec both photospheric and nebular lines appear
as broad as in other Type IIP SNe \citep[][J15]{Barbarino2014}.
This excludes the scenario of a low-energy explosion of a high-mass
progenitor, which otherwise may be able to eject silicon-layer material (note that we have not adressed constraints from the dynamics here). An upper
limit to the explosion energy for SN\,2012ec can also be deduced; if
$E$ was significantly greater than 1 B no progenitor could eject
the silicon layer and still produce as little \ni~as 0.03~\msun.

\subsection{Model grids}

\begin{figure}
\centering
\includegraphics[width=1\linewidth]{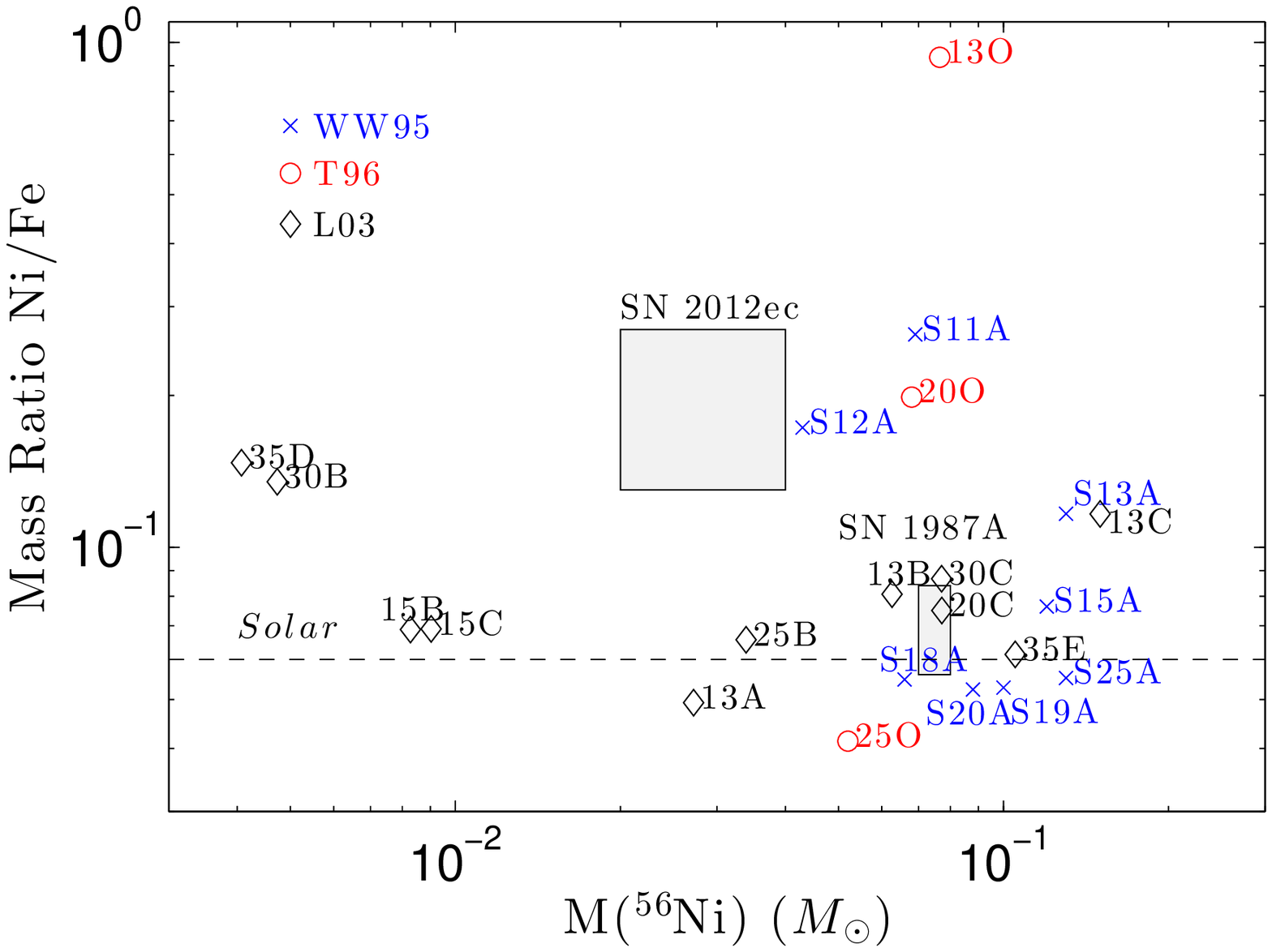} % nickelplots.m
\caption{Position of SN\,2012ec ($M(^{56}\mbox{Ni}) = 0.02-0.04$ \msun, $M(\mbox{Ni})/M(\mbox{Fe}) = 0.13-0.27$) in the 
$M(\mbox{\ni})$ - $M(\mbox{Ni})/M(\mbox{Fe})$ plane (shaded region),
compared to the model grids presented by \citet{Woosley1995} (WW95), \citet{Thielemann1996} (T96), and \citet{Limongi2003} (L03). Also shown is the location of SN 1987A.}
\label{fig:plane}
\end{figure}

The next step is to see what Ni/Fe ratios are produced in full
explosion simulations. We let $M(\mbox{Ni})$ and $M(\mbox{Fe})$
denote the mass of all Ni and Fe, respectively, one year after
explosion. SN 2012ec has $M(\mbox{Ni})/M(\mbox{Fe}) = 0.13 - 0.27$ and $M(^{56}\mbox{Ni}) = 0.02-0.04$ \msun~(J15). Figure \ref{fig:plane} shows the measured position of
SN\,2012ec in the $M(\mbox{\ni})$-$M(\mbox{Ni})/M(\mbox{Fe})$ plane
compared to the explosion simulations by \citet{Woosley1995}, T96, and \citet{Limongi2003}.

The mass of ejected nickel in these models depends on the choice of
mass cut/piston mass coordinate, which is set manually. Most of the
models in the \citet{Woosley1995} and T96 grids eject more \ni~than
produced in SN\,2012ec (so the mass cut is set ``too deep''). They also
produce a lower Ni/Fe ratio. Some of the models in the \citet{Limongi2003} grid produce lower amounts of \ni, but none has the right combination of \ni~mass and Ni/Fe ratio.

%\subsubsection{T96 model grid}

The T96 grid uses deeper mass cuts than is appropriate for SN\,2012ec. For
% TEMP unclear how mass cuts are chosen in T96
example, the mass cut used for their 20 \msun~model is 1.61 \msun,
which adds 0.02 \msun~of silicon-shell material with $\eta=0.494$ to
the ejected ashes. This gives a dramatically higher Ni/Fe~ratio than
if the mass cut would have been placed at 1.67 \msun, as needed for an
ejected \ni~mass of 0.03 \msun.
% DC 2015-05 Mcut=1.61 Msun for 20 Msun model.

For the 13 \msun~model the Ni/Fe value is 0.93. This is burning of
silicon-shell material with $Y_e = 0.491$. Comparing with
Fig.~\ref{fig:yemag}, the entropy must have been higher
than what would have been needed to make a ratio consistent with SN
2012ec. Indeed, inspection of our MESA simulation shows that the
progenitor density at the outer edge of the silicon shell is
$\rho_{\rm pre}=\mbox{few}\times10^6$ g cm$^{-3}$ so the post-shock density (which is about 7 times larger) is of order 
$\rho_{\rm post}=10^7$ g cm$^{-3}$. For $T \sim 5\e{9}$ K the entropies 
will be higher than needed for $Y_e=0.490$ in Fig.~\ref{fig:yemag}.  For
the 25 \msun~star, the Ni/Fe ratio is 0.04, too low for SN\,2012ec. The
burning occurred at lower entropy than needed for $Y_e=0.499$.
% TEMP check density in MESA model

%\subsubsection{WW95 model grid}

The WW95 explosions use a piston rather than a thermal bomb. With this
method, the choice of mass coordinate for the piston combined with a choice of explosion energy
determines the mass cut (which equals the piston mass coordinate plus
fallback). In the WW95 models the piston is placed at the edge of the
iron core.

\begin{figure}
\centering
\includegraphics[width=1\linewidth]{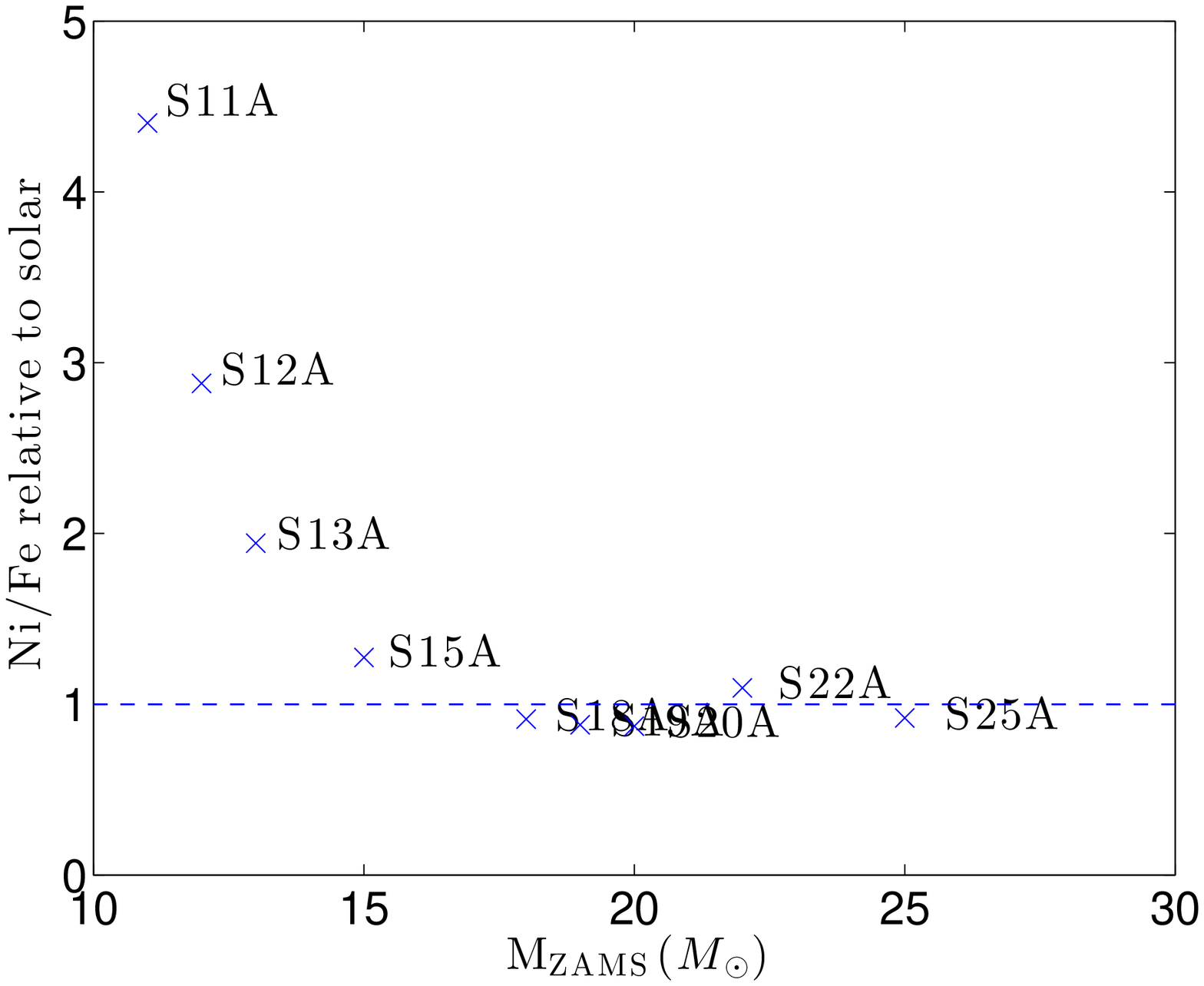} % nickelplots.m fig 2
\caption{The Ni/Fe ratio (relative to the solar value) in the WW95 model grid, for $E=1.2$\,B and solar metallicity.}
\label{fig:WW95}
\end{figure}

Figure \ref{fig:WW95} shows the Ni/Fe ratio in units of the solar value versus $M_{\rm ZAMS}$ in
the WW95 grid. The mass cut is inside the silicon-shell for 
$M_{\rm ZAMS} \lesssim 15$ \msun, raising the Ni/Fe production to several
times the solar value. This is not offset by subsolar production at
higher $M_{\rm ZAMS}$; the burning occurs on the large plateau region
in Fig.~\ref{fig:yemag}. The $M_{\rm ZAMS}=11-13$
\msun~models give high enough Ni/Fe to match SN\,2012ec, but the total
mass of the ejected ashes is somewhat too large
(Fig.~\ref{fig:plane}). The best fitting model is 12A, which makes
0.04 \msun\ of \ni~and gives a Ni/Fe ratio of 0.18, in close agreement
with SN\,2012ec.

%\paragraph{WH07 grid}
%To be done...

%\paragraph{L03 model grid}
%and L03 grids deploy mass cuts giving a variety \ni~masses.

% L03 uses Z=0.02, 13-35 Msun.

%\subsubsection{L03 model grid} 
The L03 models are piston-driven explosions and nucleosynthesis yields for different piston motions at each $M_{\rm ZAMS}$ are presented. Most models produce a Ni/Fe ratio around solar. The only models with Ni/Fe approaching SN\,2012ec (30B and 35D) only eject trace amounts of iron-group nuclei.

\section{Discussion}
\label{sec:discussion}

%\subsection{Position of the mass cut}

%The transition from the silicon shell to the oxygen shell is usually accompanied by a density jump; this facilitates shock propagation and the mass cut in self-consistently calculated models often found here (WW95, WH07). \textbf{How to get mass cut inside si shell, bernhard, alex?}

%\subsection{Asymmetries}

\subsection{Multidimensional effects}

The ordered onion-like structure of iron-silicon-oxygen shells
obtained in one-dimensional stellar evolution models may not be a good
approximation of the true structure. Multidimensional simulations of the advanced stages of
burning suggest strong convective overturns and a rearrangement of the
various burning ashes \citep{Bazan1998, Meakin2006, Arnett2011, meakin_2011_aa, viallet_2013_aa, Couch2015}.  
This opens up a possibility that neutron-rich material gets mixed out and resides at a larger mass coordinate than in 1D models, providing an
alternative means of making a large Ni/Fe ratio. Investigation of the
explosive silicon burning process in such multidimensional progenitors
would be of significant interest.

% DC 2015-05 Bazan and Arnett 1998 : 2D simulation of shell O burning.

%\subsection{Aspherical explosions}

Also the explosion may involve asymmetries. Direct evidence for asymmetries in core-collapse explosions are available from imaging of Cas A \citep{Fesen2006, Hwang2012,Isensee2010, grefenstette_2014_aa} and SN\,1987A \citep{Wang2002, kjaer_2010_aa, Larsson2013, Boggs2015}. If the explosion is asymmetric, the entropy for a fixed explosion
energy becomes higher in the direction of the explosion, and the outer
mass coordinate for complete silicon burning increases in this
direction. At the same time, we expect deeper layers to be more easily
ejected in the direction where the energy is focused. Thus, one
expects the expelled silicon burning ashes to originate from a more
extended mass range in the direction of the explosion.

\citet{Nagataki1997} studied the influence of asymmetries on the
\nif/\ni~production using two-dimensional hydrodynamic
simulations \citep[see also][]{Nagataki1998, Nagataki2000}.
The study employed two different mass cuts, the first spherically
symmetric, and the second sorting test particles by energy and making
the mass cut at the energy giving the specified amount of \ni.
For the second (more realistic) case, higher degree of explosion asymmetry gave deeper mass cuts in the explosion direction, as expected. Silicon-layer material can thus more easily be ejected in the explosion direction, thereby achieving a high Ni/Fe ratio.

Of significance for the analysis here is also their Case B model in which $Y_e$ was
artificially set to $Y_e=0.499$ everywhere. This model illustrates the sole
effect of a higher entropy achieved in asymmetric explosions ($S_\gamma$ increases by a factor of 2-3 from the spherical symmetric case to their most asymmetric model). From Fig. \ref{fig:yemag} we expect a quite modest change in Ni/Fe for an entropy change of that order; indeed the ratio varies by less than 30\% in the \citet{Nagataki1997} simulations. It is of interest to note that the ratio they obtained \emph{decreases} with higher degree of asymmetry. Considering Fig. \ref{fig:yemag} again, this is likely because the burning region spans both the $\alpha$-rich freeze-out regime (where Ni/Fe increases) and the $\alpha$p-regime (where Ni/Fe decreases), and the net effect is a small decrease. This, in fact, adds further weight to the argument that the burning in SNe giving several times solar Ni/Fe ratios cannot have occured at as high $Y_e$ as 0.499.

\subsection{Neutrino-processed ejecta}
% Neutrinos reset Ye for inner few 1e-3 Msun of ejecta.
% Depending on expandion times scales and neutrinos luminosities,
%Ye can go both up and down. But if M(56Ni) >> 1e-3 msun this
% should not be a key component
Potentially, the material with high neutron excess could also come
from the hot, neutrino-heated bubble deeper in the supernova core
(whose neutron-to-proton ratio has been reset by $n(\nu_e,e^-)p$,
$p(\bar{\nu}_e,e^+)n$, and $p(e^-,\nu_e)n$ reactions) instead of being
directly ejected after undergoing explosive burning. However, the
similarity of $\nu_e$ and $\bar{\nu}_e$ luminosities and mean energies
seen in modern simulations generally drives $Y_e$ in the innermost ejecta
above $0.5$ \citep{Froehlich2006}. Neutron-rich
conditions can only be maintained in some of the early neutrino-heated
ejecta if they expand rapidly enough, as in the electron capture
supernova model of \citet{Wanajo2011}.
% Frohlich gets Ye > 0.50 for the innermost 0.011 Msun ejecta

For SNe producing $M(^{56}\mbox{Ni})\gg 0.01$ \msun, this process does not provide a likely explanation, however. The total mass of iron group material originating from the
fast ejection of neutron-rich, neutrino-processed material is limited
to a fraction of the mass of the gain region at shock revival in this
scenario, i.e.\ to a few times $10^{-3} M_\odot$ (as in
\citealt{Wanajo2011}). Moreover, the fast ejection
of the early neutrino-heated ejecta in electron capture supernovae
depends crucially on the special density structure of their super-AGB
progenitors, and it is doubtful whether sufficiently short expansion
time-scales could be reached in more massive progenitors.

\subsection{Nickel isotopes in the Crab}
The Crab has a measured Ni/Fe ratio of $60-75$ times solar \citep{Macalpine1989}. This extreme ratio can be produced in the electron capture scenario described above, and indeed nucleosynthesis and expansion dynamics are consistent with such an origin \citep{Nomoto1982, Kitaura2006, Wanajo2009}.  It is of interest to ask whether such a process is a unique solution, or whether explosive burning without any strong neutrino processing of the ejecta may also explain this value.

Inspection of Fig.~\ref{fig:yemag}  shows that for $Y_e \ge 0.495$ (the lowest value encountered outside the iron core), the only burning conditions that give such a Ni/Fe ratio lie in the high entropy regime that is not matched by any physical shock speeds (Sect. \ref{sec:shockspeeds}). The most neutron-rich fuel $Y_e= 0.490$ can give such a Ni/Fe ratio at lower and more typical explosion entropies ($\log (S_\gamma/R) \sim +0.5$), but as discussed in Sect. \ref{sec:preSNevol}, such a $Y_e$ is too low for the silicon shell, and would be part of the iron core. It is difficult to achieve mass cuts that deep, and many other isotopes are produced in extremely non-solar proportions. It therefore appears a contrived scenario to explain the Ni/Fe ratio in the Crab without the neutrino processing occurring in an electron capture event.%, and as long as the derived value holds, can indeed be taken as evidence favoring this hypothesis for the origin of the Crab.}

% DC 2015-05 : Ni/Fe > 60 indeed requires entropies fown at rho ~ 5.5
% entropu + 0.5 ok.

\subsection{Nickel isotopes in SN\,1987A}
\label{sec:87a}

In J15 it was demonstrated that taking the optical depth of [Ni II] 6.636 $\mu$m into account, the Ni/Fe ratio in SN\,1987A derived from nebular-phase line luminosities is around solar, suggesting no or small ejection of silicon-shell material. This is consistent with the analysis in \citet{Kozma1998-II}, where a model with a nickel mass of 1.3 times solar relative to iron gave reasonable fits for [Ni~II] 6.636 $\mu$m and [Ni~II] 10.68 $\mu$m. These results supersede initial estimates of Ni/Fe $\lesssim 0.5$ times solar based on analytical formulae assuming optically thin emission \citep{Rank1988,Wooden1993}. %model luminosities in [Ni II] 6.636 $\mu$m and [Ni II] 10.68 $\mu$m lines for a model with a stable nickel mass of $6\e{-3}$ \msun. This nickel mass gave a good reproduction of the [Ni II] 10.68 $\mu$m line, which depends only on mass and temperature. The model showed some discrepancies with [Ni II] 6.636 $\mu$m, likely as a result of a too large volume assumed for the iron/nickel region.
% Kf98 uses f(Fe) = 0.55 with Vcore = 2000 km/s. Unclear why they underproduce
% Ni 6.6 mu by a factor of two. Ni
The abundance of $^{57}$Ni could be directly determined from gamma-ray lines, giving a $^{57}$Ni/\ni~ratio of 1$-$2 times the solar $^{57}$Fe/$^{56}$Fe ratio\footnote{($^{57}$Fe/$^{56}$Fe)$_{\odot}$=0.023
 \citep{Lodders2003}}\citep{Kurfess1992}. A similar range was inferred from infrared lines \citep{Varani1990} and light curve models \citep{Fransson1993,Fransson2002,seitenzahl_2014_aa}.
% DC 2015-05 Kurfess 1992 has (1-2) times solar Fe57/Fe56.

\begin{figure}
\includegraphics[width=1\linewidth]{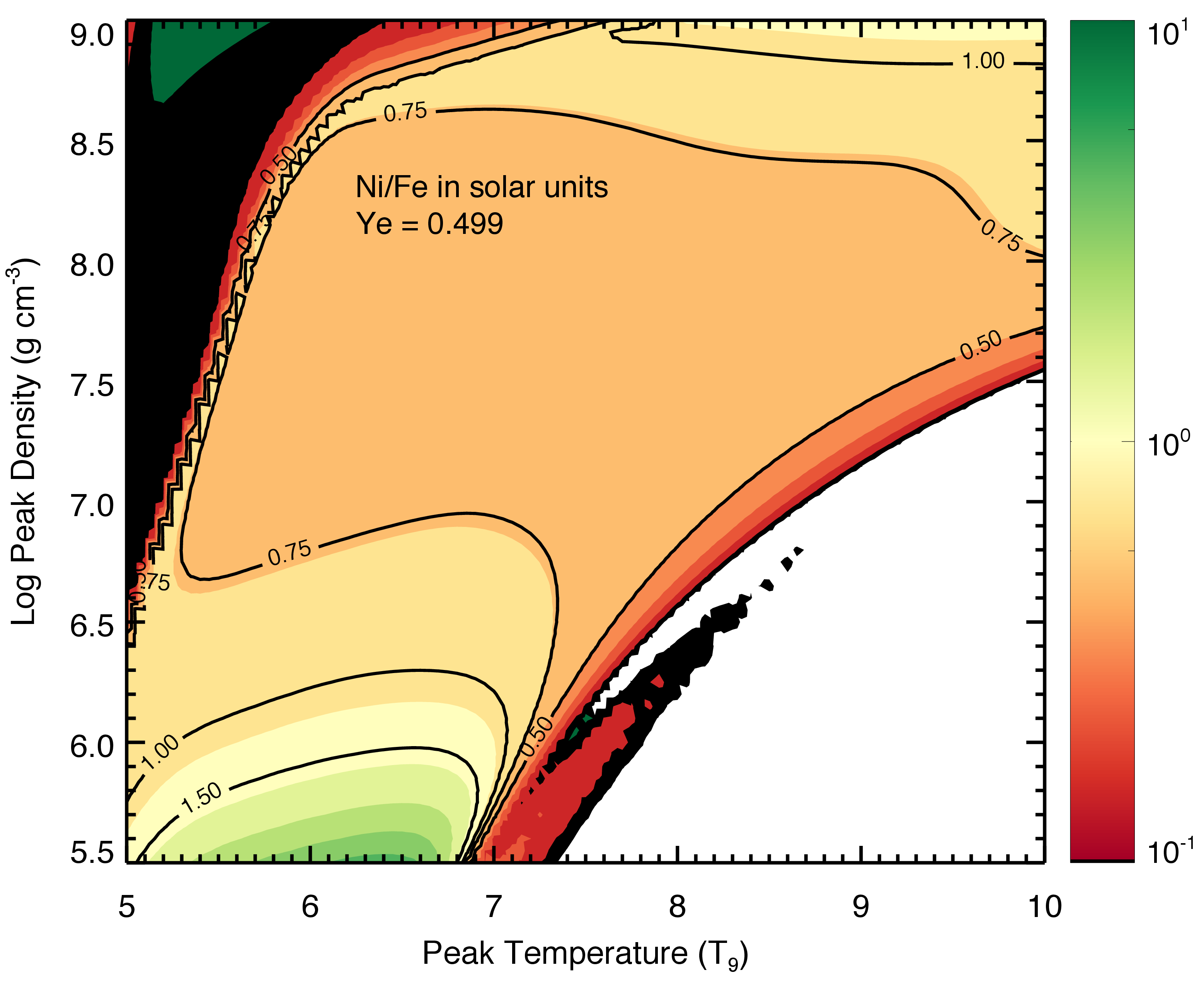}
\includegraphics[width=1\linewidth]{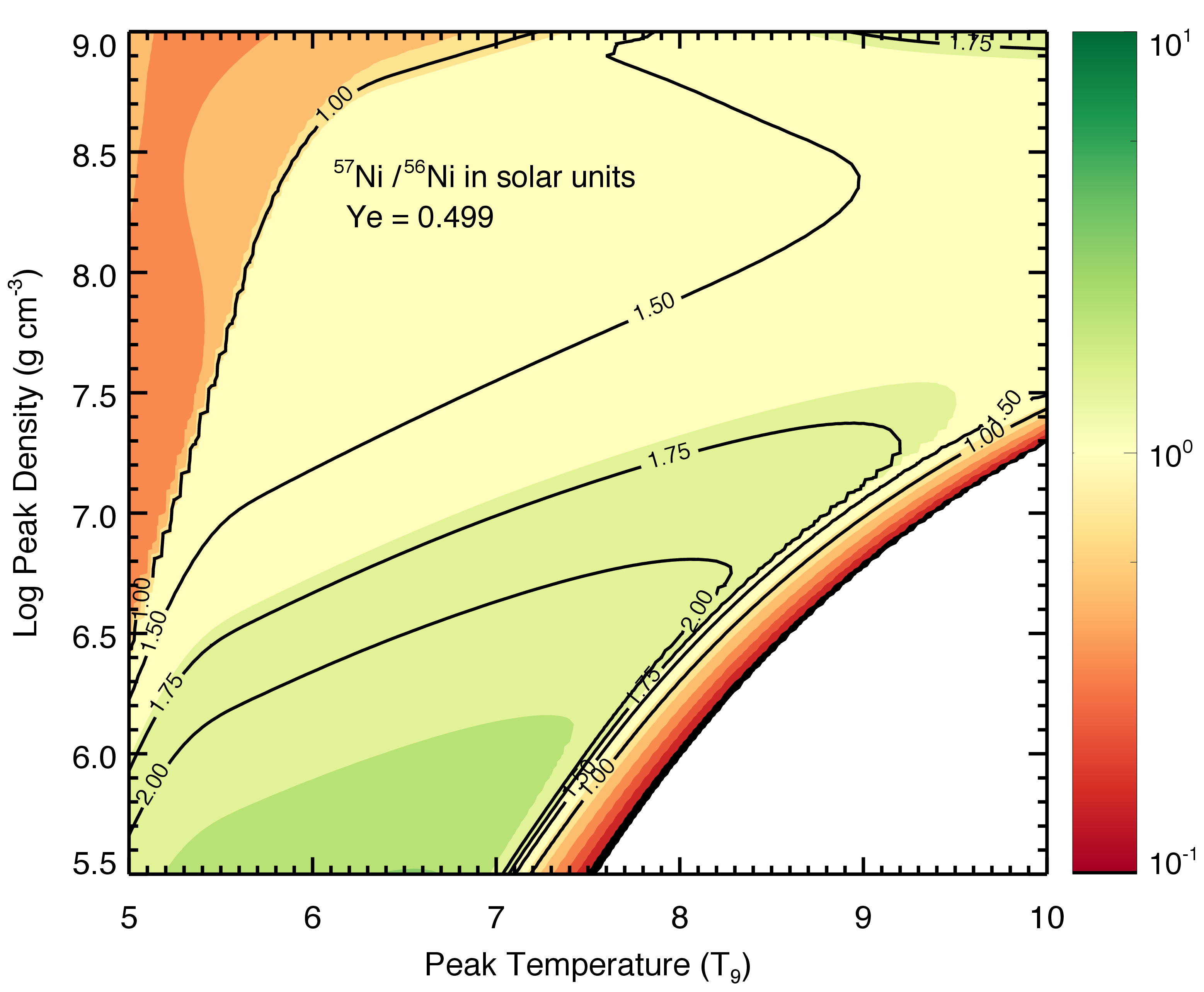}
\caption{The Ni/Fe ratio in units of the solar ratio (top) and the $^{57}$Ni/\ni~ratio in units of the solar $^{57}$Fe/$^{56}$Fe ratio (bottom), from the $Y_e = 0.499$ simulation.}
\label{fig:aaa}
\end{figure}

Figure \ref{fig:aaa} shows the production of Ni/Fe (in units of solar) and $^{57}$Ni/\ni~(in units of solar $^{57}$Fe/$^{56}$Fe) for
the $Y_e=0.499$ case from the parameterized thermodynamic trajectories of
\S \ref{sec:param}.  A Ni/Fe ratio of $0.5-1.5$ times solar and a
$^{57}$Ni/\ni~ratio of $1-2$ times solar is reproduced for burning at
post-shock densities of $\rho_{p}=10^6-10^8$ g cm$^{-3}$, which are typical values encountered in simulations (Fig. \ref{fig:yemag}). The revision of the
Ni/Fe ratio in J15 due to the calculated optical depths in the [Ni II]
6.636 $\mu$m line is therefore important in allowing a consistent
solution for the production of nickel isotopes in SN\,1987A.

A lower $Y_e$ moves the fit region to higher peak densities. For
example, $Y_e = 0.497$ would require $\rho_p > 10^8$ g cm$^{-3}$ to produce
the SN\,1987A values (not plotted, but see also \citet{Woosley1991}). This density is larger than obtained in any
published explosion models.  Thus, the most consistent picture is
that explosive burning in SN\,1987A occurred in the
oxygen shell, and that the mass cut therefore was outside the silicon
shell.
For a $M_{\rm ZAMS} = 20$\,\msun~spherically symmetric progenitor model 
this is at $\sim$1.65~\msun~in the T96 models and at $\sim$1.76\,\msun~in the WH07 models. From
this we expect the minimum value of the mass cut to be around 1.7\,\msun.
The baryonic neutron star mass was constrained to $1.2-1.7$\,\msun~from the neutrino burst \citep{Burrows1988}. This is marginally consistent with the constraints from the silicon burning, and disfavors a progenitor mass larger than 20 \msun~(but allows for smaller).
%\msun. %To eject 0.075 \msun~of \ni with zero delay time requires an explosion energy of 0.8 B.  Longer delay times allow for smaller explosion energies.
% DC 2015-01-19 : IN T97 there is an outer Si shell 1.63 - 1.67 Msun but with almost Ye = 0.499..pure O outside 1.67. In WH07 O shell starts at 1.76

\section{Summary}
\label{sec:summary}

We have analyzed the production of nickel and iron (after radioactive decays) in explosive silicon burning, and the constraints on progenitor structure
and burning conditions derivable from measured yields.  We have
focussed on the implications of Ni/Fe ratios of several times the
solar value, as recently reported for the Type IIP SN\,2012ec (J15), and previously for the Type Ic SN SN 2006aj \citep{Maeda2007, Mazzali2007}.

Using a nucleosynthesis network with parameterized thermodynamic trajectories, we have computed the
nickel and iron yields (after decays) as function of peak temperature, peak density,
and neutron excess $\eta$ of the fuel.  The Ni/Fe ratio is for most regimes
dominated by the production ratio of \nif/($^{54}$Fe + \ni). In the
normal freeze-out regime, this ratio grows with increasing entropy as
the neutron excess storage switches from $^{54}$Fe to $^{58}$Ni. When
this switch is complete, the ratio reaches a plateau value of
$\mbox{Ni}/\mbox{Fe}= \eta/0.034$; fuels with $\eta=0.034\times \mbox{Ni}/\mbox{Fe}$ provide a large region of allowed thermodynamic
conditions. Smaller values of $\eta$ can also achieve the production, but only
in a limited region of strong $\alpha$-rich freeze-out where the
$^{58}$Ni/\ni~ratio grows due to efficient depletion of \ni. We find that this process is not
likely to be responsible for producing Ni/Fe ratios several times solar as the necessary entropies would require unphysical shock speeds. Higher values of $\eta$ can achieve the production, but only in restricted regions of low entropy that are not encountered in typical explosion simulations. We conclude that to produce a Ni/Fe ratio of a few times solar, burning at normal freeze-out, or moderate $\alpha$-rich freeze-out (helium mass fraction $X(^4\mbox{He}) \lesssim 0.2$), of a fuel with $\eta \sim 6\e{-3}$ ($Y_e \sim 0.497$) is required.
% Thielemann 1990, page 227, He mass fraction reaches 0.18 in SS explosion of 87A

The derived neutron excess value can be linked to the location in the star where explosive
silicon burning occurred. Models for the progenitor structure show that the
neutron excess successively increases inwards from the oxygen shell
($\eta \sim 2\e{-3}$) to the iron core ($\eta \gtrsim 4\e{-2}$). Only in the
silicon shell is the neutron excess in the range needed for a Ni/Fe production of a few times the solar value $(\eta \sim 6\e{-3}$). An exception is if the metallicity of the star exceeds five times solar, in which case the neutron excess in the oxygen shell increases to $\eta \sim 6\e{-3}$. Such a metallicity can be ruled out for SN\,2012ec from HII region spectroscopy \citep{Ramya2007}. SN 2012ec, and other CCSNe producing large Ni/Fe ratios, therefore likely burnt and ejected
part of their silicon layers, which is a key constraint for explosion
models. %Supernovae which produce Ni/Fe ratios close to solar, such as SN\,1987A, on the other hand must have burnt and ejected only oxygen-shell material.

In spherical symmetry,  a given progenitor structure, explosion energy, delay time, and measured amount of \ni, defines the position of the mass cut. Lower-mass stars have relatively thick silicon shells that more easily encompass the mass cut. We find that 
$M_{\rm ZAMS} \lesssim 13$ \msun~progenitors exploding with a delay time of
less than one second are able to eject part of their silicon layers. Such a progenitor for SN\,2012ec is
in agreement with the estimate of $M_{\rm ZAMS}$ from modelling of the
hydrostatic burning ashes ($M_{\rm ZAMS}=13-15$ \msun, J15). Higher-mass progenitors
only eject oxygen shell material with moderate $\alpha$-rich
freeze-out, giving a Ni/Fe ratio close to the solar value. In
particular, the measured amounts of \ni, $^{57}$Ni, and
$^{58}$Ni in SN\,1987A appear consistent with the burning and ejection of
pure oxygen-shell material. This in turn translates to a lower limit for
the mass cut of this SN of 1.7 \msun.

Asymmetry in the explosion can qualitatively change the Ni/Fe ratio by
two means; by more easily ejecting deeper-lying silicon layers
in the direction of the explosion, and by achieving a stronger
$\alpha$-rich freeze-out. Published simulations \citep{Nagataki1997} show
that the entropy effect is insufficient, but the first mechanism can
achieve a high Ni/Fe ratio. An asymmetric explosion is a plausible
explanation for the high Ni/Fe yield in SN\,2006aj, an X-ray flash SN \citep{Mazzali2006, Maeda2007, Mazzali2007}.

For the Crab, a very high Ni/Fe ratio of 60-75 yimes solar has been reported \citep{Macalpine1989}. We find that this extreme value cannot be reproduced by any realistic-entropy burning outside the iron core, and neutrino-neutronization obtained in electron-capture models \citep{Wanajo2009} remains the only viable explanation.

In conclusion, it is clear that constraints on both progenitor structures and explosion dynamics can be obtained from silicon-burning yields determined from nebular-phase spectra. The exact location of the silicon burning (i.e. the mass cut) has strong ramifications for iron-group yields obtained in galactic chemical evolution models, and the constraints derived from SN~2012ec and several other CCSNe provide important input for such modeling. Further observations and modelling of nebular-phase SNe, combined with nucleosynthesis modelling, will enable us to make further progress in understanding the elusive SN explosion mechanism and the origin of the iron-group elements.

\acknowledgements
We thank K. Nomoto, F.-K. Thielemann, I. Seitenzahl, S. Rosswog, and R. Sethuram for discussion. The research leading to these results has received funding from the European Research Council under the European Union's Seventh Framework Programme (FP7/2007-2013)/ERC Grant agreement n$^{\rm o}$ [291222]. SJS acknowledges funding from STFC grants ST/I001123/1 and ST/L000709/1. FXT was supported by NASA under TCAN grant NNX14AB53G,
by NSF under SI$^2$ grant 1339600, and by NSF under grant PHY-1430152
for the Physics Frontier Center JINA. AH was supported by an ARC Future Fellowship (FT120100363) any JINA. BM acknowledges support by the Australian Research Council through a Discovery Early Career Research Award (DE150101145).

% apj
\bibliographystyle{apj}
\bibliography{bibl_part2}

\end{document}